\documentclass[12pt,preprint]{aastex}
\shorttitle{Morphology of bright galaxies in the Coma cluster}
\shortauthors{Aguerri et al.}

\begin{document}


\title{Environmental Effects in the Structural Parameters of
Galaxies in the Coma Cluster}


\author{J. A. L. Aguerri}
\affil{Instituto de Astrof\'{\i}sica de Canarias, C/ V\'{\i}a L\'actea s/n, 38205 La Laguna, Spain}
\email{jalfonso@ll.iac.es}

\author{J. Iglesias-Paramo}
\affil{Laboratoire d'Astrophysique de Marseille, BP8, Traverse du Siphon, 13376 Marseille, France}
\email{jorge.iglesias@oamp.fr}

\author{J. M. V\'{\i}lchez}
\affil{Instituto de Astrof\'{\i}sica de Andaluc\'{\i}a, Consejo Superior de Investigaciones 
Cient\'{\i}ficas, Camino Bajo de Hu\'etor 24, Apdo. 3004, 18080 Granada, Spain}
\email{jvm@iaa.es}

\and 

\author{C. Mu\~noz-Tu\~n\'on}
\affil{Instituto de Astrof\'{\i}sica de Canarias, V\'{\i}a L\'actea s/n, 38205 La Laguna. Spain.}
\email{cmt@ll.iac.es}


%\altaffiltext{1}{Visiting Astronomer, Cerro Tololo Inter-American Observatory%.
%CTIO is operated by AURA, Inc.\ under contract to the National Science
%\altaffiltext{2}{Society of Fellows, Harvard University.}
%\altaffiltext{3}{present address: Center for Astrophysics,
%    60 Garden Street, Cambridge, MA 02138}
%\altaffiltext{4}{Visiting Programmer, Space Telescope Science Institute}
%\altaffiltext{5}{Patron, Alonso's Bar and Grill}


\begin{abstract}
We have studied  116 bright galaxies from the Coma cluster  brighter than $m_{r}=17$ mag. From a quantitative morphological analysis we find  that the scales of the disks are smaller than those of field spiral galaxies. There is a  correlation between the scale of the disks and the position of the galaxy in the cluster;  no large disks are present near  the center of the cluster or in high density environments. The structural parameters of the bulges are not affected by the environment. We have analyzed the distribution of blue and red objects in the cluster.  For spirals there is a trend between color and position in the cluster.  The bluest spiral galaxies are located at larger projected radii; they also show larger velocity dispersions than the red ones. The differences in the scale of the disks between cluster galaxies and local samples of isolated galaxies and the color distribution of the objects can be understood in terms of the harassment scenario.
\end{abstract}




\section{Introduction}

The effect of  environment on the evolution of galaxies remains a fundamental question. The environment of clusters provides an ideal laboratory for testing the effects of galaxy density, galaxy interactions, and interaction with the intracluster medium on the properties of galaxies. In this respect the analysis of quantitative morphology of cluster galaxies can provide useful evolutionary constraints. There  are two main theories for explaining  galaxy evolution in clusters. The model proposed by Merritt (1984) states that the morphology distribution of the galaxies does not change after the first cluster collapse. This implies that, excluding the core region, the tidal effects are not important in the morphology of the galaxies. During recent years high resolution numerical simulations have shown the evolution of galaxies inside a cluster-like potential. The scenario proposed by Moore et al. (1996, 1998, 1999) concludes that the evolution of galaxies in clusters is driven by tidal interactions with the cluster gravitational potential and high speed galaxy--galaxy encounters. This harassment mechanism  can be active even until late  the evolution of
the cluster. It has also been proposed that effects caused by the strong pressure shocks over the interstellar medium produced by the displacement of the galaxies through the hot intracluster medium (Gunn \& Gott 1972) can strip the HI reserves and could be partly responsible for the morphological evolution observed in clusters (Dressler et al 1997).

Two possible observational tests of the harassment mechanism are the morphological distribution of the galaxies in the cluster and the enhancement of the star formation rate (SFR) due to tidal forces with the gravitational potential and galaxy--galaxy encounters.  

From the pioneering work of Dressler (1980), it was observed that the morphology of the galaxies in clusters depends on the position in the cluster (see Thomas 2002 and reference therein). Late-type galaxies are located in the outermost parts (low density regions of the cluster), while early-type objects are in the central and higher density cluster regions. In the harassment scenario the morphology of the galaxies changes on short time scales when the
galaxies are falling from the outermost parts of the cluster into the central regions (Gunn \& Gott 1972; Charlot \& Silk 1994; Abraham et al. 1996; Dressler et al. 1997; Ellis et al. 1997; Oemler et al. 1997). This mechanism is very efficient at transforming late-type galaxies into dwarf spheroidals and early-types into S0s, and at reducing the size of the disks. The role of the cluster environment in galaxy evolution has been analyzed in the past by studying different aspects of galaxies: asymmetries in the rotation curves of galaxies in clusters (Rubin et al. 1988; Whitmore et al. 1988; Dale et al. 2001; Dale \& Uson 2003), the HI deficiency of cluster spirals (Cayatte et al. 1990; Haynes et al. 1984; Solanes et al. 2001), and observations of head--tail radio sources and ram pressure stripped spirals (Kenney \& Koopman 1999). 

Galaxy clusters at medium redshift ($z=0.2$--0.5) show an excess of blue objects compared with clusters at the present epoch (Kodama \& Bower 2001). This is the so-called Butcher--Oemler effect (Butcher \& Oemler 1978). Several mechanism have been proposed in order to explain this effect. Fujita (1998) demonstrates that the star-forming activity can be enhanced because of the tidal force from the potential  of the cluster, or high speed galaxy--galaxy encounters. He also predicted a different distribution of the blue galaxies in the cluster depending on the responsible mechanism of the star formation. Blue galaxies have also been observed in nearby clusters (McIntosh et al. 2003), but in smaller numbers than at medium redshifts. They are located at the outermost regions of the clusters and are consistent with galaxies undergoing the first harassment when they are falling  into the cluster center. 

The aim of this paper is to perform a quantitative analysis of the distribution of the morphology and colors of a large sample of bright galaxies in the Coma cluster in order to compare the structural parameters of Coma galaxies and isolated local samples. We will also study the distribution of dwarfs and bright galaxies in the cluster. This will tell us about the role of the cluster environment in  galaxy evolution. Coma is a rich and nearby cluster which has been analyzed extensively in the past. It is one of the brightest X-ray emission clusters (White et al. 1993). Dynamical studies confirmed that this is not a relaxed cluster (Fitchett \& Webster 1987; Baier et al. 1990). Coma ellipticals have been used is several studies of the fundamental plane (Mobasher et al. 1999; Scodeggio et al. 1998; Khosroshahi et al. 2000). Morphological studies have also been made (Lucey et al. 1991; Jorgensen \& Franx 1994; Andreon et al. 1996, 1997; Gerbal et al. 1997; Khosroshahi et al. 2000), although most of them have been based on qualitative or visual classifications. Recently, Guti\'errez et al. (2003) made the quantitative morphology of a sample of galaxies in the core region of Coma cluster, the central 0.25 $\square^{2}$. Our study is the first to analyze extensively  quantitative morphology of the Coma galaxies for a wide field of view ($\approx 1 \square^{2}$). The advantages of the quantitative morphology are that it is reproducible and biases can be understood and carefully characterized through simulations that are treated as real data. The use of quantitative morphology allows the recovery of reliable information on the structural parameters of the galaxies. The information contained in the structural parameters plays a necessary role in the understanding of the evolution and origin of galaxies. In this paper we will compare the structural parameters of the Coma cluster galaxies with those from field.

%\section{The data}

The observations were taken with the Wide Field Camera (WFC) at the prime focus of the  2.5 m Isaac Newton Telescope (INT)  at the Roque de los Muchachos Observatory (La Palma), in  2000 April. The plate-scale of the CCD was 0.333 arcsec/pixel, and the seeing of the images was about 1.5$''$. Four exposures of the Coma cluster  were obtained under photometrical conditions, covering an area of $\approx 1\ \deg^{2}$. The images were obtained using the 
broad band Sloan--Gunn $r'$ filter of the WFC. For a detailed description of the observations, data reduction and calibration see Iglesias-P\'aramo et al. (2002, 2003).

\section{Surface brightness decomposition}

We have fitted the isophotal surface brightness profiles of the galaxies
with two components (bulge and disk). The surface brightness profiles of the
bulges were modeled by a S\`ersic (1968) law, given by

\begin{equation}
I(r)=I_{\rm e} 10^{-b_{n}((\frac{r}{r_{\rm e}})^{1/n}-1)},
\end{equation}
where $r_{\rm e}$ is the effective radius, which encloses half of the intensity
of the profile, $I_{\rm e}$ is the effective intensity, and $n$ is the profile
shape parameter. The parameter
$b_{n}$ is given by $b_{n}=0.868n-0.142$ (Caon et al. 1993). 

The disks were fitted by exponential profiles (Freeman 1970):

\begin{equation}
I(r)=I_{0} e^{-r/h},
\end{equation}
where $I_{0}$ is the central intensity and $h$ is the scale length
of the profile. A Levenberg--Marquardt nonlinear fitting algorithm
was used to determine the parameter set that minimizes $\chi^{2}$.
We used the algorithm design by Trujillo et al. (2001). This
algorithm takes into account the seeing effects on the surface
brightness profiles and the intrinsic ellipticity of the galaxies.
It was successfully applied in determining the quantitative morphology
of field galaxies (Aguerri \& Trujillo 2002) and galaxies in clusters
(Trujillo et al. 2001, 2002). Extensive Monte Carlo
simulations were carried out in order to determine the uncertainties
in the determination of the structural parameters.

\subsection{Monte Carlo Simulations}

The proper determination of the structural parameters of the objects depends mainly on the signal-to-noise ratio of the images. In order to quantify this effect, we have run  Monte Carlo simulations to determine the limiting magnitude for which the recovered parameters
are well determined. These simulations allow us to detect possible biases in the recovered parameters. We have generated galaxies with similar sizes and
luminosities to those expected in Coma. We simulated two types
of galaxies: those formed only by a bulge component (S\`ersic's law) and
others formed by a bulge and disk (S\`ersic + exponential). We have added
 random noise to the models, which was the obtained from a region
free from sources from the images (see Trujillo et al. 2001 for
a more detailed about the simulations).

The magnitudes of the simulated objects were in the interval 13 mag $\leq m_{r} \leq19$ mag. The scales of the objects simulated  by a S\`ersic profile (bulge galaxies) were 0.5 kpc $\leq r_{\rm e} \leq 10$ kpc, and $1\leq n \leq 6$. The bulges of the
bulge + disk galaxies have scales of 0.3 kpc $\leq r_{\rm e} \leq 5$ kpc and $1\leq n \leq 6$, and the scales of the disks were   $h \leq 6$ kpc. The bulge-to-total ratio, $B/T$, of the objects varies in the interval $0 < B/T \leq 1$. The parameters of
the 300 simulated objects were chosen randomly distributed on the previous intervals. In this paper we will use the value $H_{0}=75$ km s$^{-1}$ Mpc$^{-1}$.

Figure 1 shows the relative errors between the input and output
parameters for the simulations for bulge galaxies. All the parameters
are recovered with errors less than 20\% down to  $m_{r}=19$ mag. Figure 2
shows the relative errors for the parameters of the simulated
bulge + disk galaxies. Although, these simulations show larger errors in the recovered parameters, all the parameters can be obtained with
errors of less than 20\% for the objects with $m_{r}\leq 17$ mag.  

Figure 3 shows the recovered $B/T$ ratio. This parameter is recovered
with a mean difference of less than 0.1 for galaxies with $m_{r}\leq 17$
mag  and
simulated objects with input $B/T\leq 0.8$: objects with $B/T>0.8$ were recovered with a mean difference larger than 0.1. It should be noted that disks with $h<2 kpc$ are not recovered well, even in those objects with $m_{r}\leq17$ mag. They are very small and unrealistic disks. We adopted
$m_{r}=17$ mag as the limiting magnitude  for our fits.  All galaxies with $m_{r}\leq 17$ mag were decomposed in bulge and disk, those with $B/T > 0.8$ or
those not well fitted by a bulge + disk, were
fitted by a single S\`ersic profile.

\subsection{Galaxy classification}

Our images cover an area of 1 square degree of Coma cluster. We have selected in this area all galaxies with $m_{r} \leq 20$ (see Iglesias-P\'aramo et al 2003). Only those galaxies in Coma\footnote{We assumed that Coma members are those with 4000 km s$^{-1}$ $\leq cz \leq 10000$ . The radial velocities were taken from Nasa Extragalactic Database (NED).} with $m_{r}\leq 17$ (215 objects) were taken into account in the present work. They are the  86$\%$ of the galaxies in Coma with $m_{r}\leq17.0$. This will be the  completeness of our  sample.

First, we classified the galaxies by  luminosity. We
distinguish two main classes: bright and dwarf objects. Dwarf
galaxies are those with $M_{B}\geq -18$ mag,\footnote{The $B$ magnitude of the objects were obtained from Godwin et al. (1983).} and bright ones, those with
$M_{B} < -18$ mag. This classification gave 99 dwarfs and 116 bright
objects. 

The bright objects were classified into  morphological
types according to their $ B/T$ parameter, following Simien \& de Vaucouleurs (1986).
Galaxies with $B/T>0.8$ were classified as elliptical (E),
and their surface brightness profiles were fitted  by a single
S\`ersic law. The galaxies with $B/T\leq 0.8$ were classified into
four types: late-type galaxies (Spl), $0\leq B/T \leq 0.24$;
early-type (Spe), $0.24< B/T \leq 0.48$; lenticular (S0),
$0.48< B/T \leq 0.6$, and ellipticals with disk (Ed), $0.6<B/T \leq 0.8$.
The number of objects of each galaxy type are: 16 Spl, 20 Spe,
13 S0, 12 Ed, and 34 E. The remaining 21 objects were edge-on
galaxies, irregular objects for which we could not determine
 isophotes, or galaxies with prominent bars for which
we did not perform the decomposition of the surface brightness
profiles.

We have compared our classification with visual classifications available
for some of the galaxies of the sample. Figure 4 shows the comparison between
the classification based on the $B/T$ ratio and the $T$ morphological parameter
given by de Vaucouleurs et al. (1991) (hereafter RC3). The classification of
late-type galaxies coincides with both methods, but the agreement is
weaker for early-type objects, in particular for E, Ed, and  S0 galaxies. Based on this result, we will group E and Ed galaxies into one group, hereafter
called elliptical (E). We will define the S0 galaxies as a separate group. These types of objects are the transition between disk-dominated galaxies (normal spirals) and bulge-dominated objects. 

Figure 5 shows the fits of the isophotal surface brightness profiles for bulge galaxies, and Fig 6 shows the isophotal fits for bulge + disk galaxies. Andreon et al (1997) measured from the luminosity curve the effective radius of the early type galaxies brighter than $m_{B}=16.5$, assuming a mean color $B-R =1.5$ for Coma, they studied all galaxies within one degree from the center brighter than $m_{R}=15.0$. Our sample goes deeper ($m_{r}=17.0$). We have 10 early type galaxies in common with Andreon et al (1997). We have compared their effective radius of those galaxies and our effective radius determined from the fits of the surface brightness profiles. The agreement is better than 10$\%$. We have also compared the magnitudes of the galaxies calculated by the fitted models and those given by Godwin et al. (1983) (see Fig. 7). It can be seen that the agreement between the magnitudes is in all cases better than 0.5 mag. Gutierrez el al (2003) showed the quantitative morphology for Coma galaxies in the core cluster region. Although the method applied was the same, the surface brightness profiles of the galaxies were different, because they were obtained from different images. The final classification of the galaxies were also slightly different. In Gutierrez et al (2003) it was considered as E galaxy those with $B/T>0.6$, the surface brightness profile of these objects were fitted with only one S\`ersic profile. In the present work, E galaxies were objects with $B/T>0.8$. We have compared the structural parameters of the objects in common in both samples and with $B/T<0.6$ and $B/T>0.8$. They correspond to E and Spirals in both samples. For the E galaxies the mean relative errors of the structural parameters are: $\Delta \mu_{e}=0.007, \Delta r_{e}=0.043$ and $\Delta n= -0.075$. For the bulges of the spiral galaxies the mean relative errors are: $\Delta \mu_{e}=0.038, \Delta r_{e}=-0.064$ and $\Delta n= 0.083$,and for the disks: $\Delta \mu_{o}=0.052$ and $\Delta h=0.093$. These differences are smaller than the typical errors in the determination of the parameters ($\approx 10 \%$), and they come mainly due to differences in seeing or sky substraction on the images. The largest errors are in the case of the parameters of the discs, which are more affected by the sky background substraction.

\section{Morphological distribution of the galaxies in the cluster}

From the pioneering work of Dressler (1980), it is well known that there is a
morphological segregation of  galaxies in clusters. Early-type objects are
located in denser regions and are therefore closer to the center of the cluster than late-type galaxies. We have studied the morphological segregation in Coma based on  our galaxy classification. The morphological evolution of galaxies in clusters is mainly driven by global and local conditions. In order to study the global conditions in the cluster, the distribution of the different morphological types is studied as a function of the projected distance of each galaxy to the center of the cluster. Meanwhile, the local conditions are usually parameterized by a local projected density (Dressler 1980).

\subsection{Morphology--radius relation}

We have computed the projected radius of each galaxy to the center of the
cluster. The center adopted was that of Godwin et al. (1983): $\alpha$(J2000) = 12:59:42, $\delta$(J2000) = 27:58:15.6. The projected radius was scaled to the scale-length of the mass profile of the cluster: $r_{\rm s}=0.26 $ Mpc (Geller et al. 1999). Figure 8 shows the cumulative functions of the different types of galaxies vs.\  projected radius. The Kolmogorov-Smirnov probabilities for the Spe-Spl, Spe-S0 and Spl-S0 cumulative functions are: 0.67, 0.23 and 0.09, respectively. According with these probabilities, we can not disprove that the cumulative functions of Spe-Spl, Spe-S0 and Spl-S0 come from the same distribution function. The mean positions, $<R/r_{s}>$, of Spl, Spe and S0 galaxies are: 2.28, 2.07 and 1.61, respectively. So, Spls are located in average at  larger distances than Spes and S0s. The increment of E galaxies for $R/r_{\rm s}> 4$ corresponds to  a group of early-type galaxies (8 objects, mainly Es) located at Coma NE region. Dwarf galaxies show a gap in their distribution. There are no dwarfs within $R/r_{\rm s} \approx 0.3$. 

The absence of dwarf galaxies in the central region of the cluster could be related to  disruption caused by tidal interactions.  Debris of these kinds of systems has already been observed in the innermost regions of Coma (Gregg \& West 1998). In their study of the central region of Coma,  Trujillo et al. (2002) showed that dwarfs have a constant galaxy-light concentration. In their models, this was compatible with dwarfs being removed from the center of the cluster. Other authors have given other evidence to prove this result (Thompson \& Gregory 1993; Secker \& Harris 1997; Gregg \& West 1998; Adami et al. 2001; Andreon 2002).

\subsection{The morphology-density relation}

This relation takes   local effects on  morphological segregation
into account. Dressler (1980) defined the projected density 
by taking into account the ten nearest neighbors to each galaxy, $\rho_{10}$.  In Fig. 9 we show $\rho_{10}$ vs.\ the projected radius. It can be seen that there is a correlation between these two quantities. This means  that $\rho_{10}$ is giving us the same information as the projected radius, i.e., it is related to global effects in the cluster. That was also proposed by Thomas (2002). Figure 9 also shows the projected density defined with the nearest neighbors, $\rho_{1}$, vs,\ the projected radius. It is clear that these two parameters are weakly correlated: therefore, we will use $\rho_{1}$ as
an indicator of  local effects in the cluster.

Figure 10 shows the cumulative distributions of the different morphological types as a function of $\rho_{1}$. The mean density, $<log(\rho_{1})>$, of E, Spl, Spe, S0 and dwarfs are: 2.06, 1.69, 1.79, 1.94 1.95, respectively. This means that E and Spl galaxies are located at the largest and lowest local densities, respectively. The Kolmogorov-Smirnov probabilities for the Spe-Spl, Spe-S0 and Spl-S0 cumulative functions are:  0.66, 0.27 and 0.19, respectively. This means that we can not disprove that the cumulative functions of Spl-Spe, Spe-S0 and S0-Spl galaxies  come from the same distribution function. We have also computed the Kolmogorov--Smirnov probability for the S0--E, S0--dwarf, and E--dwarf cumulative functions to be 0.81, 0.99, and 0.62, respectively. This means that the S0s and dwarf galaxies are compatible with  the same cumulative distribution function. But, we can not disprove that S0--E and E--dwarf  have the same distribution function. It is also clear from Fig. 10 that no spirals (early- or late-type) are located at $\log(\rho_{1}) > 2.4$, while S0s, Es, and dwarfs are located at higher local densities.

\subsection{Density images}

We can define the surface number density of galaxies on the plane of the sky. We have followed the definition given by Trevese et al. (1992). For each point we find the distance, $R_{n}$, to the $n$th neighboring galaxy. The surface density at the point $(x,y)$ is then given by

\begin{equation}
\sigma(x,y)=\frac{C}{R_{n}^{2}},
\end{equation}
where $C$ is a constant determined by the condition that the integral of $\sigma(x,y)$ over the total area is equal to the total number of galaxies. We have taken into account ten neighboring galaxies as the definition of the density. Figure 11 shows the density image for each type of galaxy. E galaxies are located at the center of the cluster and show a very concentrated density. Dwarf galaxies are distributed in a shell surrounding the center of the cluster. S0s have two   density peaks near  the cluster center and  show a more extended density map than the Es. The density peak of the Es is located between the two density peaks of the S0 galaxies. Spirals (Spe and Spl) show density peaks far from the cluster center and  also  the most extended density maps. Late-type spirals show a peak in  density farther from the cluster center than the peak of the Spes.  An asterisk in Fig. 11 marks the center of the cluster. The 
E and Spe density peaks are close to the cluster center, and the Spl
the peak  is the most distant from the center. 

\section{Structural parameters}

In this section, we will study the structural parameters of the bright galaxies analyzed in Coma.  In particular, we will compare the disk and bulge parameters with those from local and isolated galaxies in order to explore the influence of the environment. The local samples used as comparison were: Graham (2001, 2003), de Jong (1996), and M\"ollenhoff \& Heidt (2001).  The sample of 86 spiral galaxies from Graham (2001, 2003)  covers all spiral galaxy types and provides the largest homogeneous set of structural parameters for spiral galaxies modeled with an $R^{1/n}$ bulge. We took the structural parameters in the $R$ band of these galaxies, for a direct comparison with our data. We will also compare our results with those given by de Jong (1996), hereafter DJ96, and M\"ollenhoff \& Heidt (2001), hereafter MH01. The structural parameter of the sample from DJ96 are obtained in the $B$ and $K$ bands with an assumed  exponential profile for the surface brightness of the bulge. This will make  a direct comparison with the parameters of our bulges impossible. The sample of MH01 is in the near-IR $J$, $H$,  and $K$ bands, with S\`ersic profiles used
for the bulges of the galaxies.

\subsection{Disk parameters}

Figure 12 shows the correlations of the scale length of the disks with  $B/T$,  the projected radius, and  $\rho_{1}$. There is no trend in the scale of the disk with $B/T$. Assuming that the morphological type of the objects is correlated with the $B/T$ ratio (Simien \& de Vaucouleurs 1986), we can infer that the scale of the disk is independent of the morphological type. Similar results were obtained in the samples from Graham (2001, 2003), DJ96 and MH01. Our mean value of $\log(h)$ is 0.33, which is smaller than the mean value obtained by Graham (2001, 2003): $\log(h)=0.59\pm0.03$ in the $R$ 
band.\footnote{The mean value of $\log(h)$ from the sample of Graham (2003) has been obtained taking into account the galaxies that lie in the same range of absolute magnitudes as the galaxies in our sample ($M_{B}=[-18.0,-21.3]$).} {This result is in agreement with the obtained recently by Guti\'errez et al. (2003) for the central region of the cluster ($R/r_{s} \approx 1.6$).

Figure 12 also shows the relation of the scale length of the disks and the local and global environmental parameters. We have overploted the mean $h$ of the disks from the Graham (2001, 2003) sample. It can be seen that the scales  of the disks from Coma are smaller than those in isolated galaxies. From Fig. 12 we can infer that no large disks are located at the center and high densities in the cluster.  Tables 1 and 2 show the parameters of these relations, the Pearson correlation coefficient ($r$), and the significance of the correlation $(P)$, that is, the probability that $\mid r \mid$ should be larger than its observed value in the null hypothesis. From table 1 can be seen that the strongest correlation between $h$ and projected radius is for Spl galaxies. Table 2 shows that the strongest correlation between local density and radius is when we consider all galaxies together.

\subsection{Bulge parameters}

Figure 13 shows the correlations of the bulge effective radius, $r_{\rm e}$ with the projected radius and projected density. There is no correlation between the bulge scale and the environment. This lack of correlation confirms a previous result by Andreon (1996)

Figure 14 shows the relation between the structural parameters of the bulges. There is a  correlation ($r=0.36, P=0.005$) between the S\`ersic index, $n$, and $B/T$. This correlation was found for the first time by Andredakis et al. (1995), and is in agreement with that found by MH01 ($r=0.48$). In the  ($\mu_{\rm e}$, $r_{\rm e}$) plane there is the following correlation: 
$\mu_{\rm e}=20.15 + 2.97 \log(r_{\rm e})$, with $r=0.67, P=0.007$,  $\mu_{\rm e}$ being the effective surface brightness of the S\`ersic profile. This is the so-called Kormendy relation. The sample of MH01 shows this relation with a similar slope and correlation coefficient to that of our sample. There is a strong correlation ($r=0.9, P=0.003$) between $\mu_{0}$ and $n$,  also found by MH01.  It should be noted that the different morphological types are located in different regions of the ($\mu_{0},n$) plane; bulges of early-type galaxies show brighter $\mu_{0}$ and higher values of $n$ than  late-type bulges. There is also a not statistically significant correlation ($r=0.46, P=0.07$) between $n$ and $r_{\rm e}$, which is not so strong as that showed by MH01 ($r=0.66$).

We have overplotted on Fig. 14 the results obtained by Graham (2001, 2003). The values of $\mu_{e}$, $r_{e}$ and $n$, from Graham (2001, 2003) field galaxies, are similar to the values  obtained for the bulges of Coma galaxies. In Coma galaxies there are more objects with large values of $r_{e}$ and $n$ than in the filed sample.  This is because Graham's sample is biased to late type galaxies.

All these correlations indicate that bulges of the galaxies in Coma are similar to those of local and isolated galaxies. Environmental effects have a weak role in the evolution of the bulges of the galaxies in this cluster.

\section{Color distribution of the galaxies in the cluster}

We have investigated the $B-R$ color distribution of the galaxies in the cluster.  Butcher \& Oemler (1984)  defined the fraction of blue galaxies, $f_{B}$, as the ratio of galaxies 0.2 mag bluer than the mean color of the E/S0 sequence to the total number of galaxies in the cluster. We have followed the same criterion. Figure 15 shows the  color--magnitude relation of all galaxies in Coma brighter than $m_{r}=17$ mag.  Applying this criterion for Coma,  $f_{B}=$ 0.07, which is smaller than the value observed in other medium-redshift clusters (Abraham et al. 1996; Rakos et al. 1997; Kodama \& Bower 2001). The origin of these blue populations has been explained in terms of galaxies falling into the cluster potential (Abraham et al. 1996), or an SFR enhanced by harassment interaction between galaxies (Rakos et al. 1997). 

Colles \& Dunn (1996) studied the blue galaxies of Coma and inferred that they  are falling into the cluster potential because of the different velocity dispersion, $\sigma$, between red and blue objects. We have additional information about the morphology. Figure 16 shows the $B-R$ colors of the different types of galaxies as a function of the projected radius to the center of the cluster. The colors of Es and S0s are constant with radius, and  Spes and Spls shows bluer colors at larger radii. This trend is more important for Spl galaxies. The correlation between the color of Spl galaxies and the projected radius to the cluster center is statistically significant ($r=-0.53$ and $P=0.04$). No correlation has been found for Spe galaxies ($r=-0.37$ and $P=0.11$). Figure 17 shows the number of red and blue galaxies as a function of the projected radius. The blue galaxies are located in the outer region of the cluster ($R/r_{\rm s} \approx 3$), while the red galaxies are more concentrated toward the center. It can be seen in Fig. 17 that there is a peak of red objects at $R/r_{\rm s} \approx 5.5$. This peak corresponds to the the galaxies belonging to a group which located at  the N-E border of the cluster. This group is basically formed by E galaxies.

\section{Discussion}

Galaxies in clusters are affected by their environment, which can be responsible for the morphological transformations between the different galaxy types. The main mechanisms proposed for such transformation are galaxy harassment, ram-pressure stripping, and pre-processing.

Galaxy harassment is the tidally induced evolution of galaxies
brought about by multiple high speed encounters with other massive galaxies and with the cluster gravitational potential (Moore et al. 1996, 1998, 1999). This mechanism, as shown by the simulations, is very efficient in transforming 
late-type galaxies into  dwarf ellipticals or dwarf spheroidals and early-type galaxies in S0s (Moore et al. 1999), and it  will also reduce the size of the disks. Similar to harassment but more efficient in rich galaxy clusters is the pre-processing mechanism, which consists in recent arrivals to the cluster being pre-processed in group or subcluster environments (Kodama \& Smail 2001; McIntosh et al. 2003). The infall of groups of galaxies with lower velocity dispersion into the cluster potential can produce a pre-processing in the evolution of the galaxies into clusters that would  be more efficient than harassment because of the lower velocity dispersion of the galaxies compare with the velocity dispersion of the galaxies in rich clusters. Another mechanism is  ram-pressure stripping. This is the removal of the cold, neutral gas reservoir in a star-forming spiral by its rapid motion through the hot ICM (Gunn \& Gott 1972: Quilis et al. 2000).

Ram-pressure stripping  can explain the color distribution of the galaxies in some clusters, but  does not imply the loss of stars. For this reason, this mechanism cannot explain the small scale of the disks that we have observed in Coma cluster galaxies. Cluster galaxies have high speed encounters and interaction with the galaxy potential during   infall into the cluster. These processes strip stars from the disks of the galaxies and scatter them into the intracluster region forming part of the intracluster light. Galaxy harassment 
and pre-processing can produce this effect. In the numerical simulations by Moore et al. (1996, 1998, 1999) it is observed that galaxy harassment makes the disk smaller. This can explain the smaller size of the disks of Coma,  compared with the field, galaxies. The pre-processing of the galaxies in groups can also reduce the scale of the disks. Colless \& Dunn (1996) conclude that Coma shows a group of galaxies around NGC 4839 merging with the two cDs located in the central part of the cluster.  This group is outside the field of our images, and we cannot see the effects of pre-processing from this group in our galaxies. But we observe  an excess of early-type (E, S0 and Spe) galaxies at a distance of $R/r_{\rm s} \approx 5.5$. This group is located at the NE border,and is formed by 8 red objects (see Fig. 17) mainly  E galaxies. It shows two disk galaxies located at $R/r_{\rm s} \approx 5.5$, which have small disks, as  can be seen  on Figure 12. They show small values of $h$  according to their position in the cluster (see Fig. 12). This could be an indication that these objects have undergone  pre-processing  in this group.

The colors of the galaxies also depend on the environment. Star formation in the galaxies can be enhanced or suppressed by  environmental conditions. Galaxy harassment includes tidal interaction of the galaxies with the global potential of the  cluster and galaxy--galaxy encounters. These two effects can be distinguished by studying the SFR of the galaxies (Fujita 1998), which can be analyzed by studying the color distribution of the galaxies in the cluster. Thus, if the cluster tidal effects enhance the SFR in the galaxies, then the bluest galaxies should be located near the center of the cluster, whereas they should be in the outer parts of the cluster if the SFR is induced by  high speed galaxy--galaxy encounters (Fujita 1998). As we showed in the previous section, the blue galaxies in Coma are located in the outer parts of the cluster. This distribution of blue and red galaxies is to be expected if the SFR is enhanced by galaxy--galaxy harassment encounters (Fujita 1998). This result has also been
observed in other galaxy clusters (Abraham et al. 1996; Rakos et al. 1997). The color variation in galaxies can be explained by differences in metallicity. Using the stellar population models from Vazdekis et al. (1996) and assuming an IMF with $\mu=1.35$, a variation in metallicity from [Fe/H] = $-0.4$ to [Fe/H] = 0.4 gives  $\Delta(B-R) \approx 0.3$, independent of the age. Previous results (Carter et al 2002; Mehlert et al 2003; Moore et al 2002; Odell et al 2002; Poggianti et al 2001) have discovered a gradient in metallicity with cluster distance for the Coma galaxies. In particular, Carter et al (2002) obtained a difference of -0.2 in [Fe/H], using Vazdekis et al (1996) models this gives $\Delta(B-R) \approx 0.11$. But, assuming  solar metallicity, the change in $B-R$ due to an age variation of 11 Gyr is $\approx$0.6. A variation in metallicity, then,
 cannot explain the variation in color observed in Spls. We also know that the blue objects show different velocity dispersions from those of the red ones (Colless \& Dunn 1996). This can be explained by their falling into the cluster. They also have the largest scale of the disks. The mean value of $log(h)$ for the disks of the bluest spiral galaxies ($B-R\leq1.5$) is: $log(h)=0.57$, which is bigger than the mean value of $log(h)$ for the galaxies in the cluster and similar to the value of the field galaxies. This is compatible,
then, with the infall of objects into the cluster undergoing the harassment for the first time.

\section{Conclusions}

We have obtained the structural morphological parameters of a sample of 116 bright galaxies in the Coma cluster. The main results are:

\begin{itemize}
\item The scale length of the disks of the galaxies are independent of the
morphological type of the galaxy as in field samples.
\item The scale length of the disks are smaller than those of field galaxies. There is a  correlation of the scales of the disks with the projected radius and the local density. This means that no large disks are observed close to the center of the cluster or in high density environments. The galaxy harassment scenario can  explain this.
\item The structural parameters of the bulges show similar relations to those of local isolated galaxies. No correlation is found between the structural parameters of the bulges and the environment.
\item The blue population of galaxies is smaller than in other clusters at medium redshift ($f_{B}=0.07$). The color of E and S0 galaxies is constant with the radius. Spe and Spl galaxies show the bluest color for those objects located at larger projected radii and show a larger velocity dispersion than red ones. This behavior is  clearer for Spl galaxies. The bluest objects are also those with the biggest disk scales. This is compatible with the picture in which the  blue objects are those falling into the cluster potential and those which are undergoing the harassment for the first time.
\end{itemize}

\clearpage

\begin{figure}
\plotone{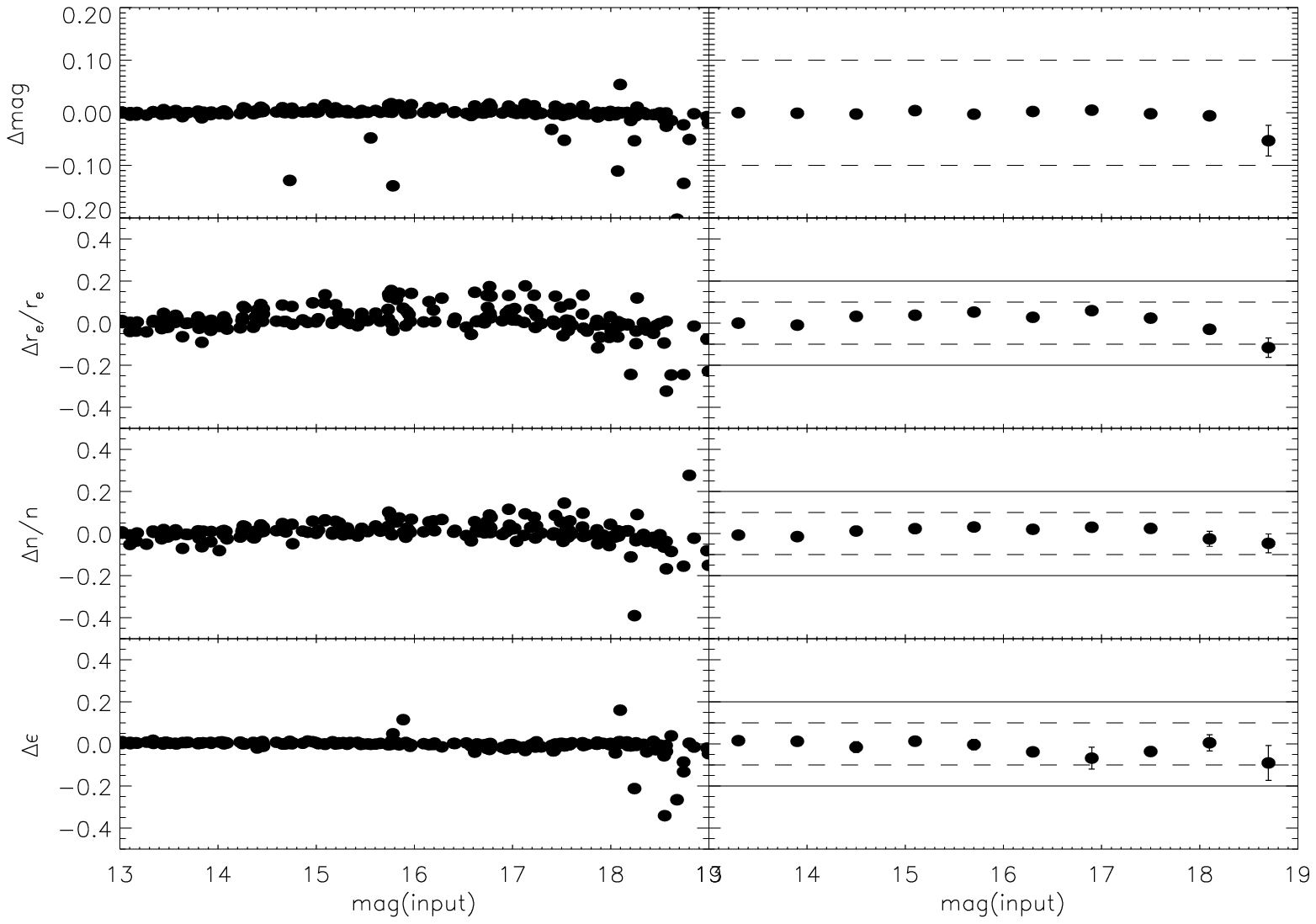}
\caption{Relative errors in the bulge parameters from Monte Carlo simulations of bulge galaxies (left panels). Mean relative errors in the bulge parameters showed in the left panels with 1$\sigma$ error bars (right panels). Horizontal dashed and full lines represent 10$\%$ and 20$\%$ relative errors of the $r_{e}$ and $n$ parameters, respectively. They also represent a difference of 0.1 (dashed line) and 0.2 (fulll line) between the input and recovered magnitud and ellipticity of the models. \label{fig1}}
\end{figure}

\clearpage

\begin{figure}
\plotone{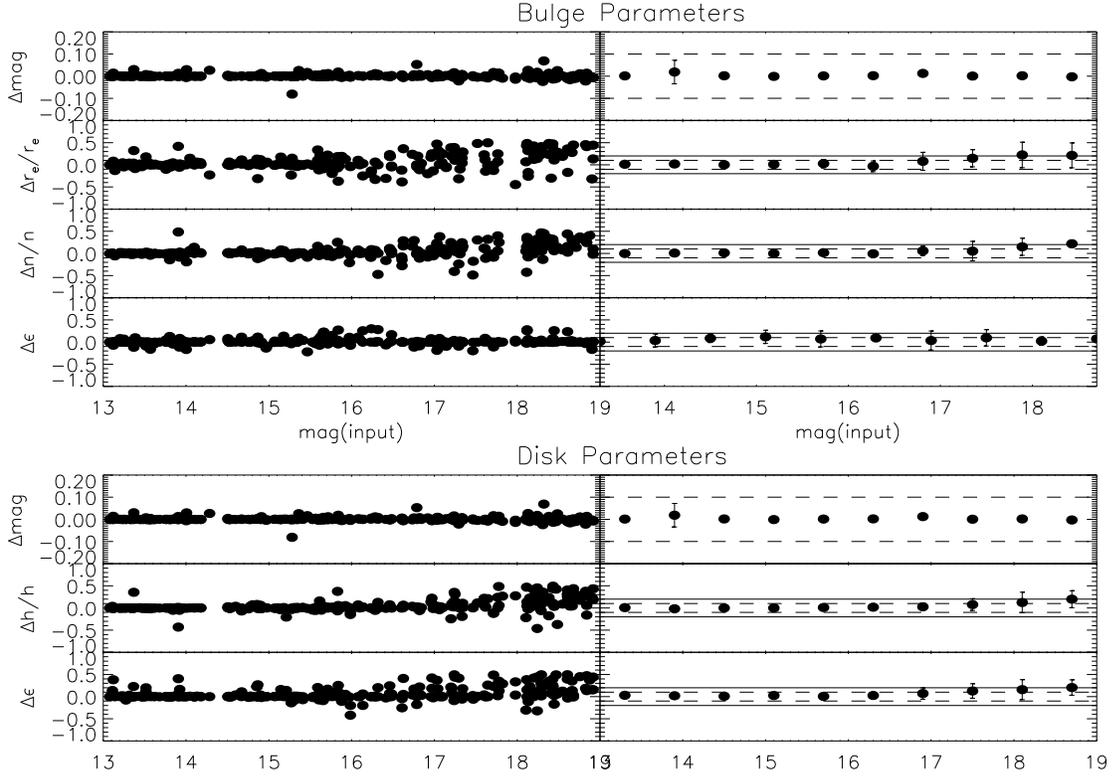}
\caption{Relative errors in the bulge and disk parameters from Monte Carlo simulations of bulge + disk galaxies (left panels). Mean relative errors in the bulge and disk parameters showed in the left panels with 1$\sigma$ error bars (right panels). Horizontal dashed and full lines represent 10$\%$ and 20$\%$ relative errors of the $r_{e}$, $n$ and $h$ parameters, respectively. They also represent a difference of 0.1 (dashed line) and 0.2 (fulll line) between the input and recovered magnitud and ellipticity of the models. \label{fig2}}
\end{figure}

\clearpage

\begin{figure}
\plotone{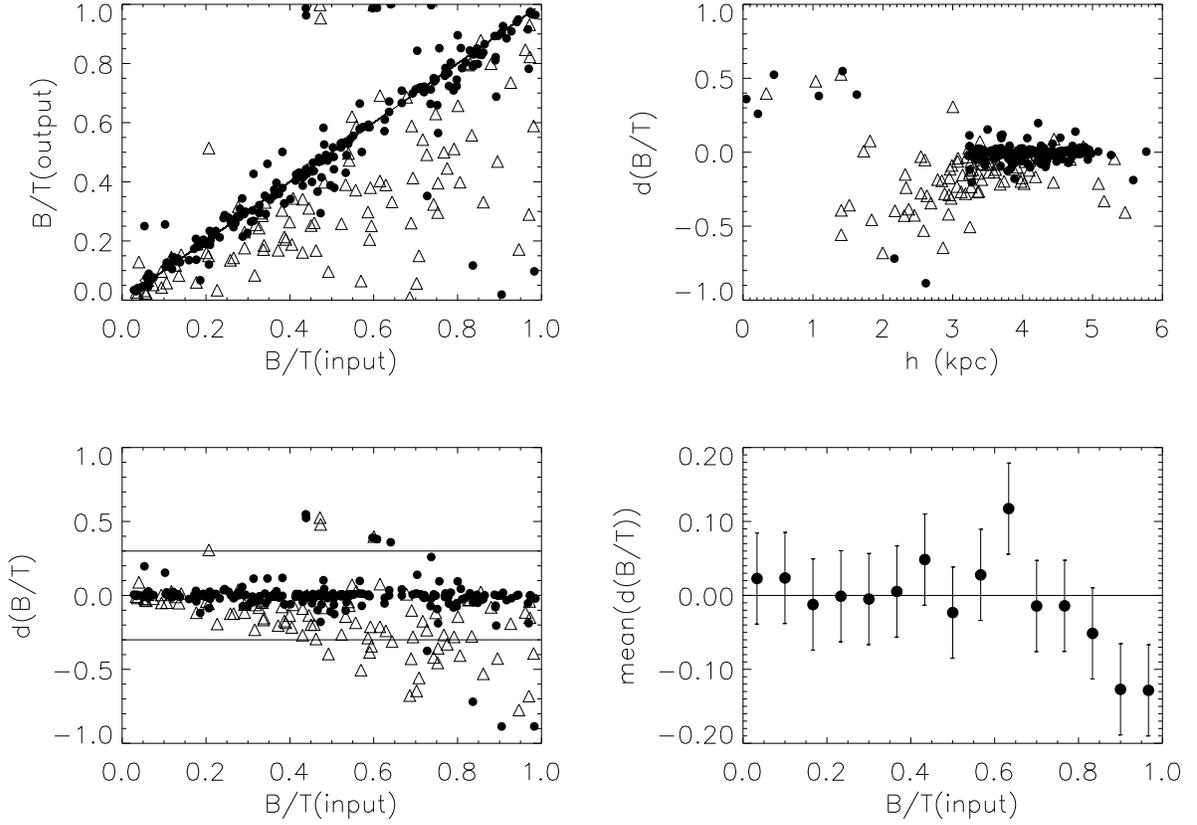}
\caption{Errors in the $B/T$ parameter from the Monte Carlo simulations. Triangles represents galaxies with $m_{r} > 17$ mag, and full circles objects with $m_{r}\leq 17$ mag.\label{fig3}}
\end{figure}

\clearpage

\begin{figure}
\plotone{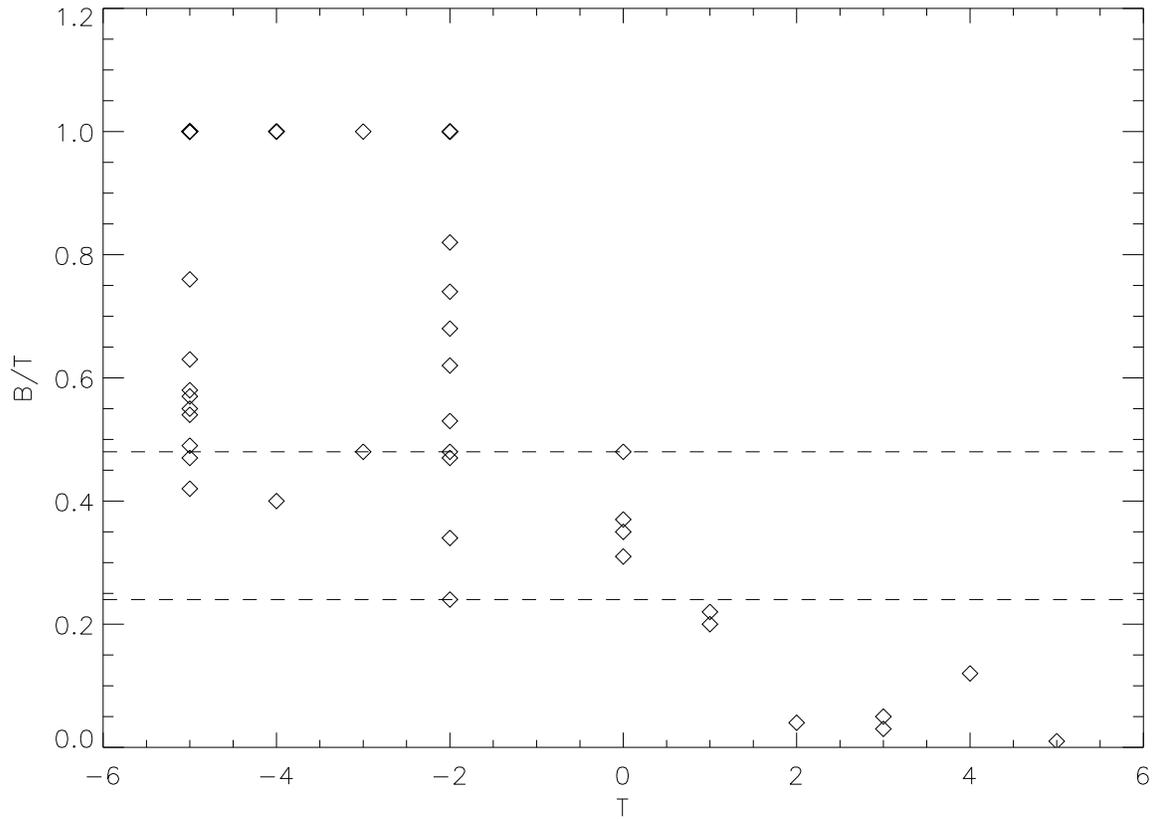}
\caption{The classification parameter $T$ from RC3 versus the $B/T$ ratio of some objects in the sample . The two horizontal dashed lines correspond to $B/T=0.24$ and 0.48. These are the limits used in this study for separating  Spl--Spe and Spe--S0 (see text for more details).\label{fig4}}
\end{figure}

\clearpage

\begin{figure}
\plotone{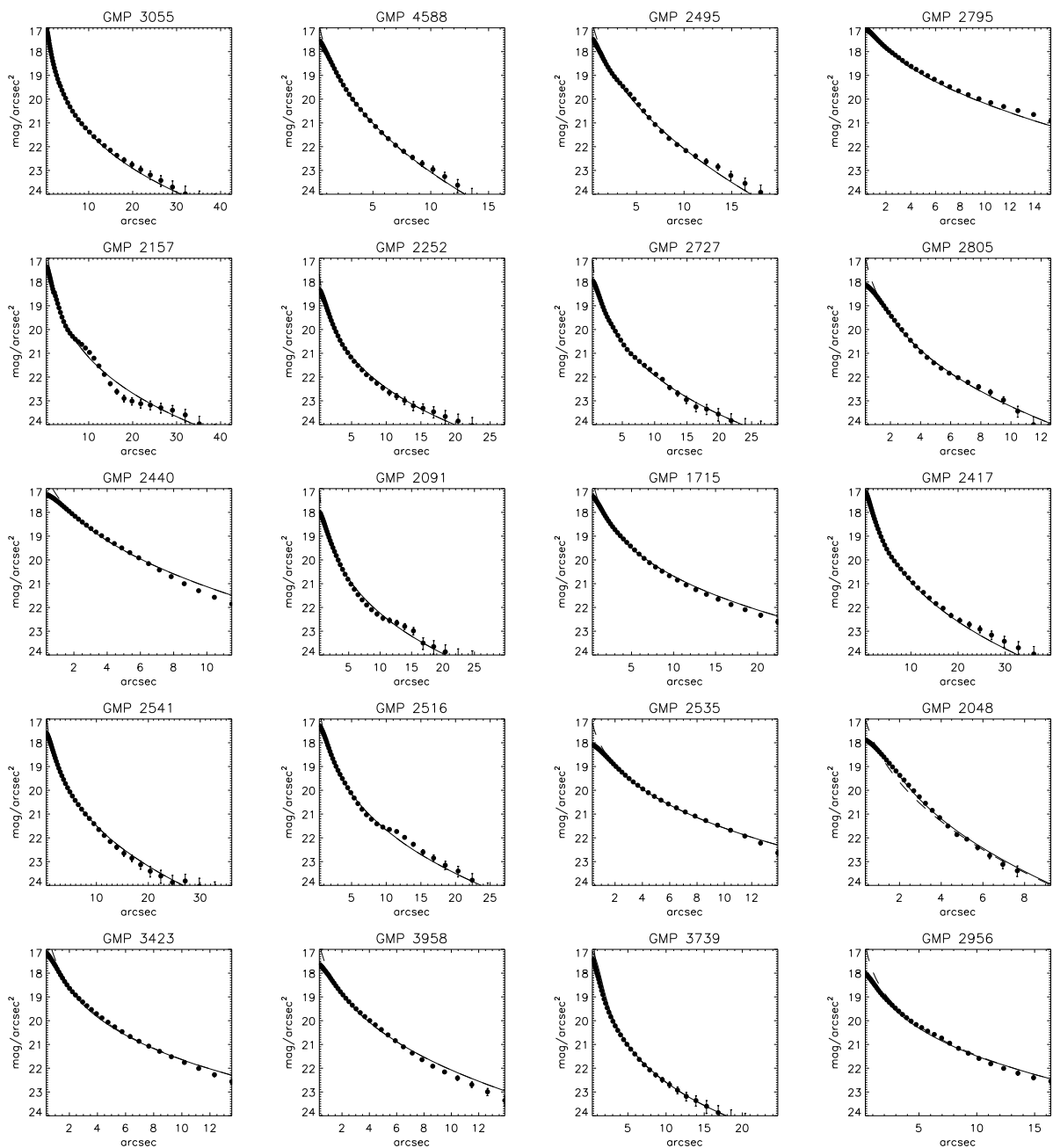}
\caption{Fit of the surface brightness profiles of the galaxies classified as
E. Only a S\`ersic profile was fitted to all profiles. The
unconvolved (dashed line) and convolved (full line) fitted S\`ersic profile
are superimposed.}
\end{figure}

\addtocounter{figure}{-1}

\clearpage

\begin{figure}
\plotone{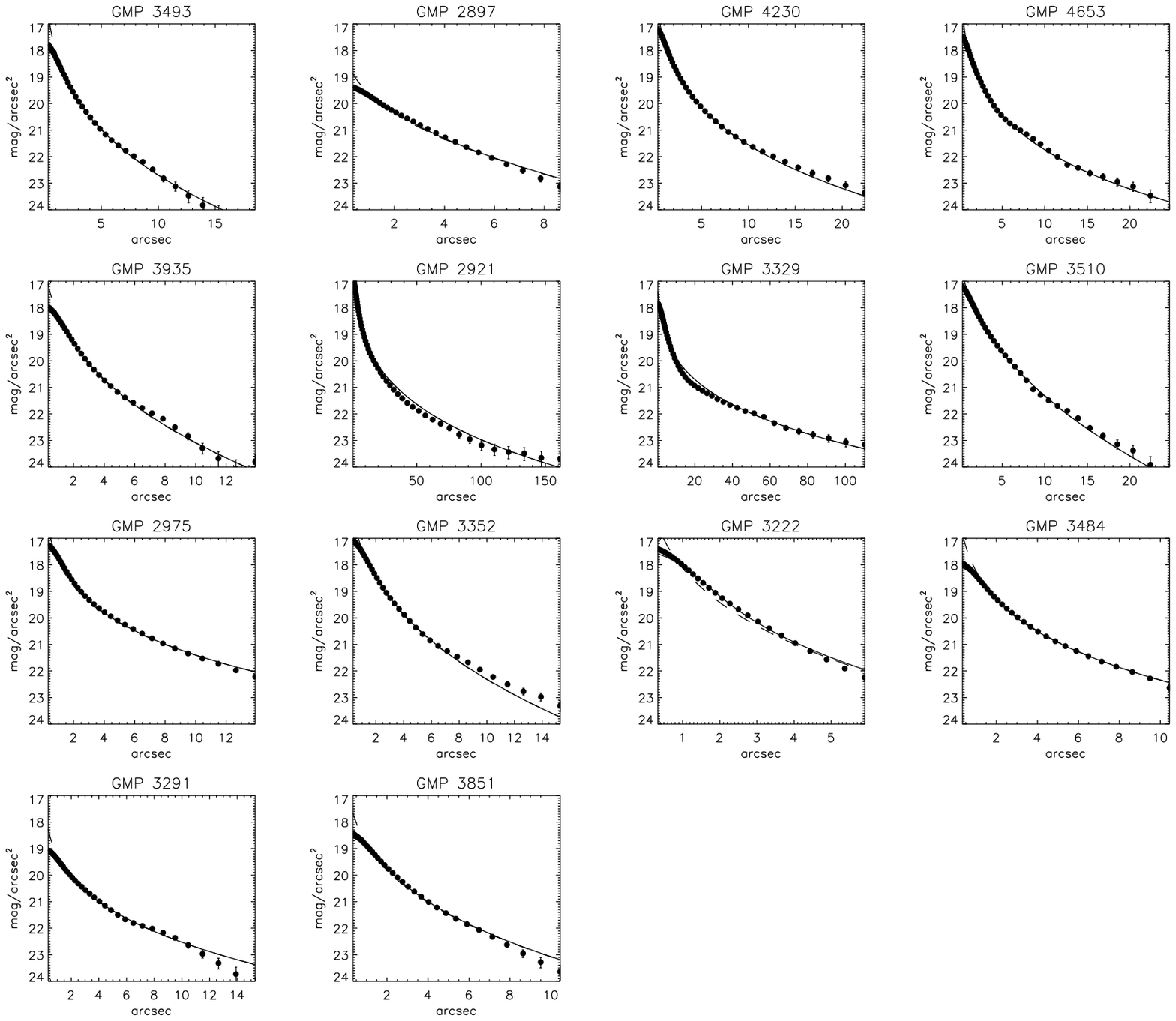}
\caption{Continue}
\end{figure}

\clearpage

\begin{figure}
\plotone{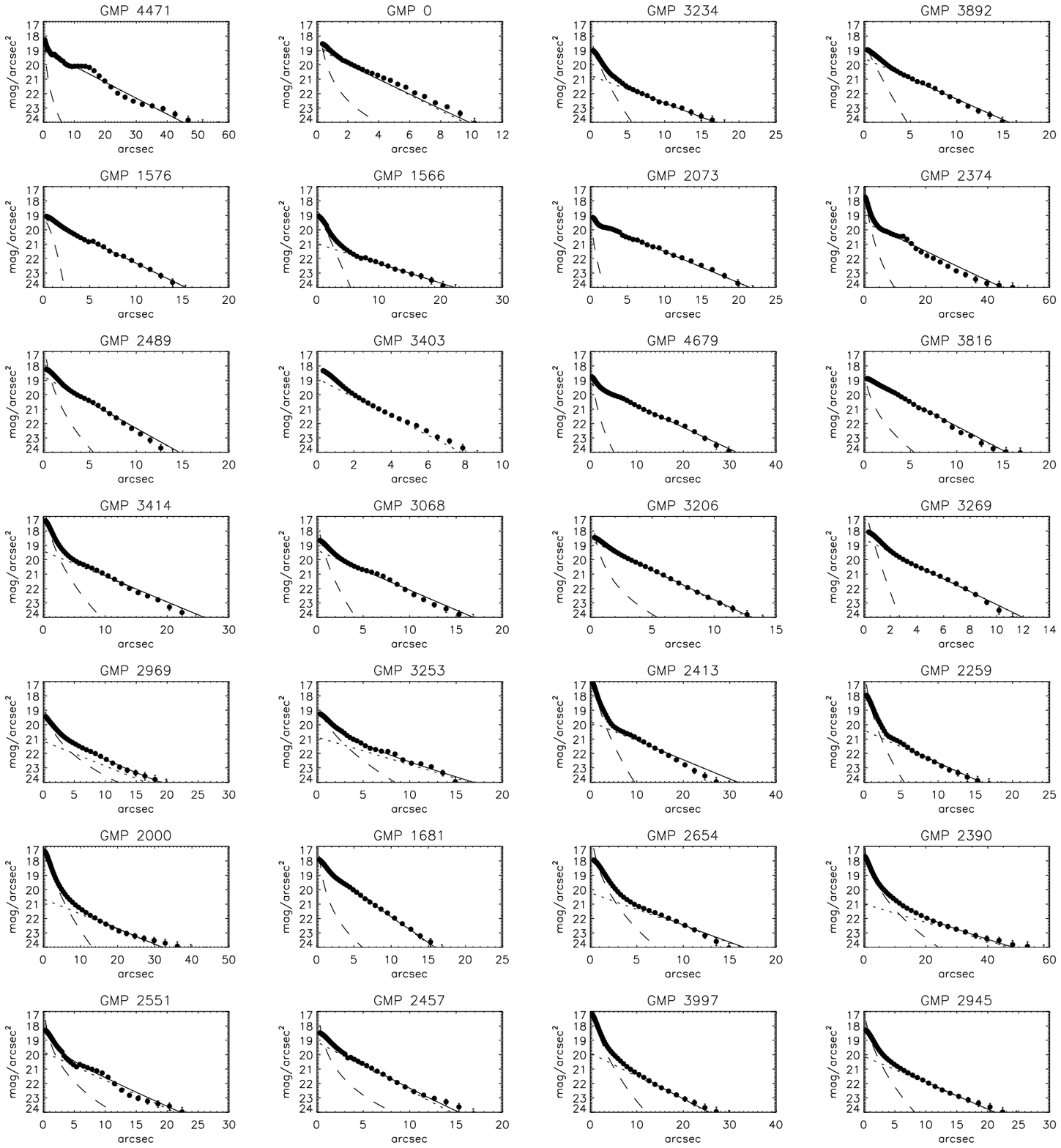}
\caption{Fits of bulge (dashed line)  and disk (dotted line) profiles to the surfaces brightness of the
galaxies. The total model (bulge + disk) is represented by the full line.}
\end{figure}

\addtocounter{figure}{-1}

\clearpage

\begin{figure}
\plotone{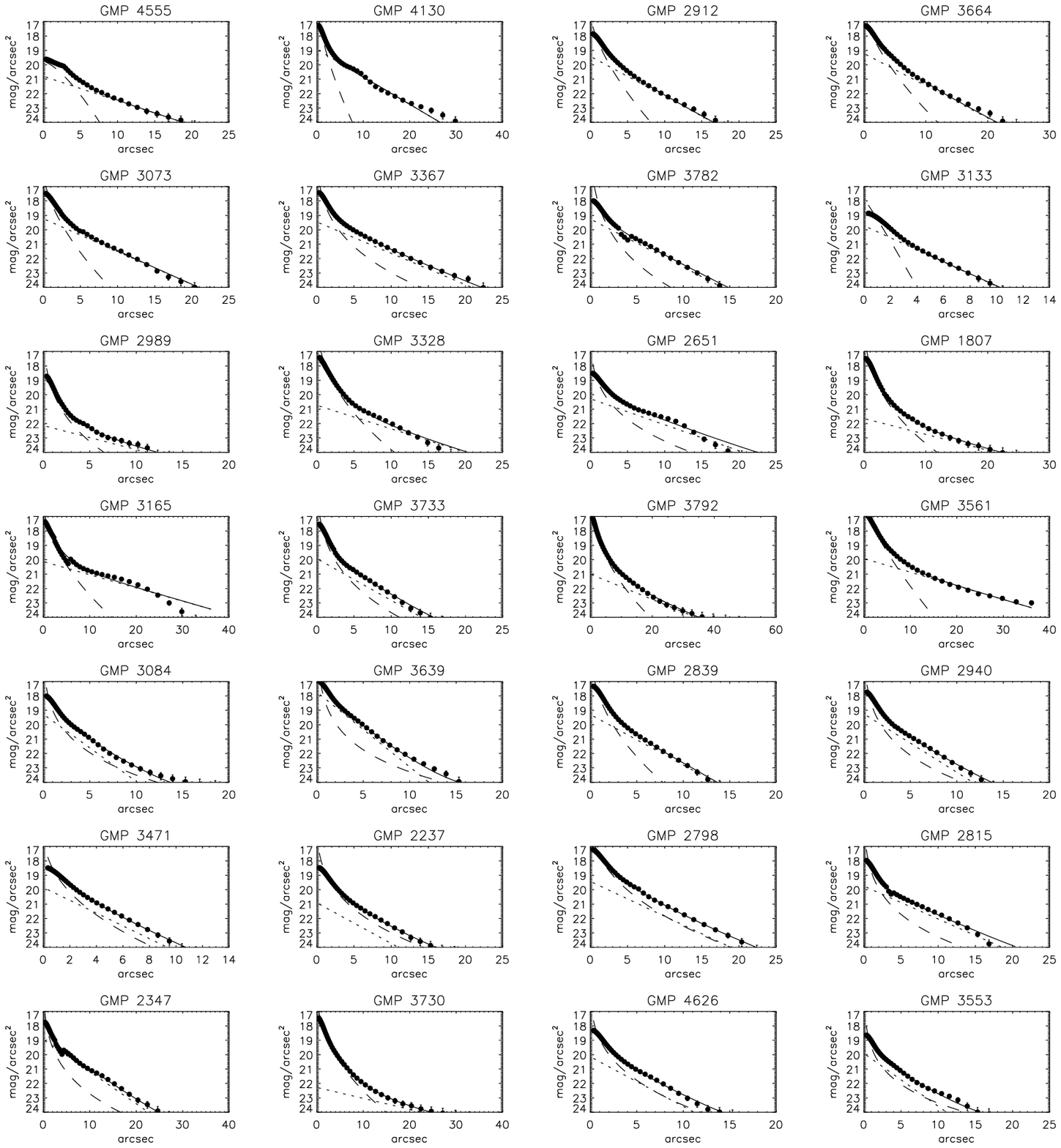}
\caption{Continue}
\end{figure}

\addtocounter{figure}{-1}

\clearpage

\begin{figure}
\plotone{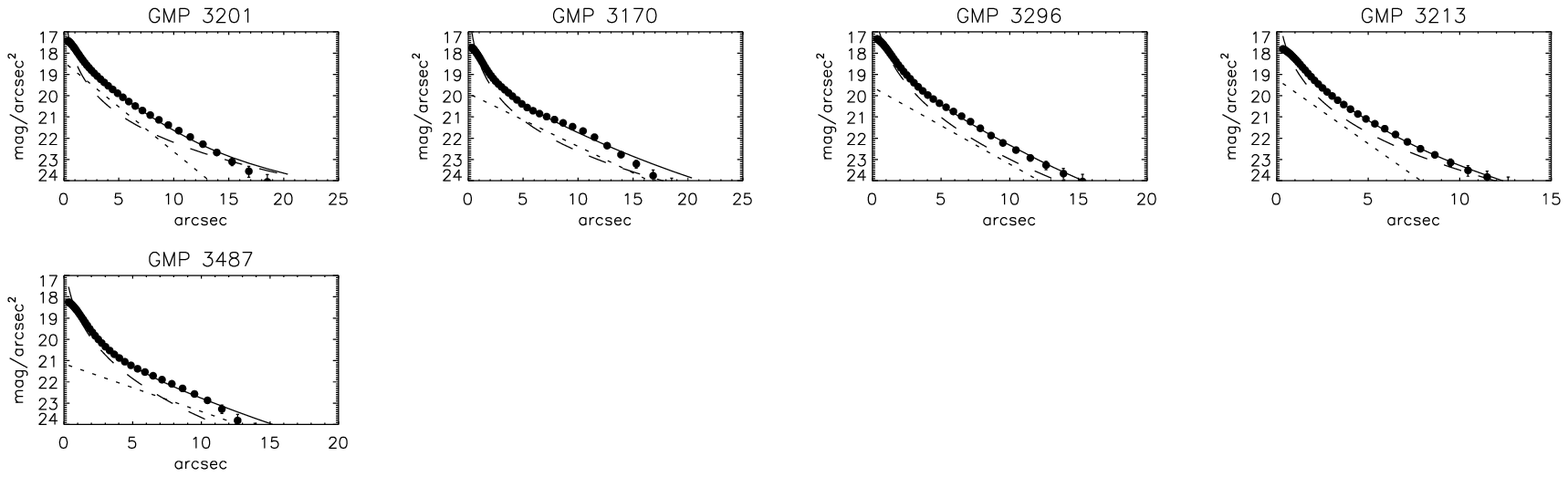}
\caption{Continue}
\end{figure}

\clearpage

\begin{figure}
\plotone{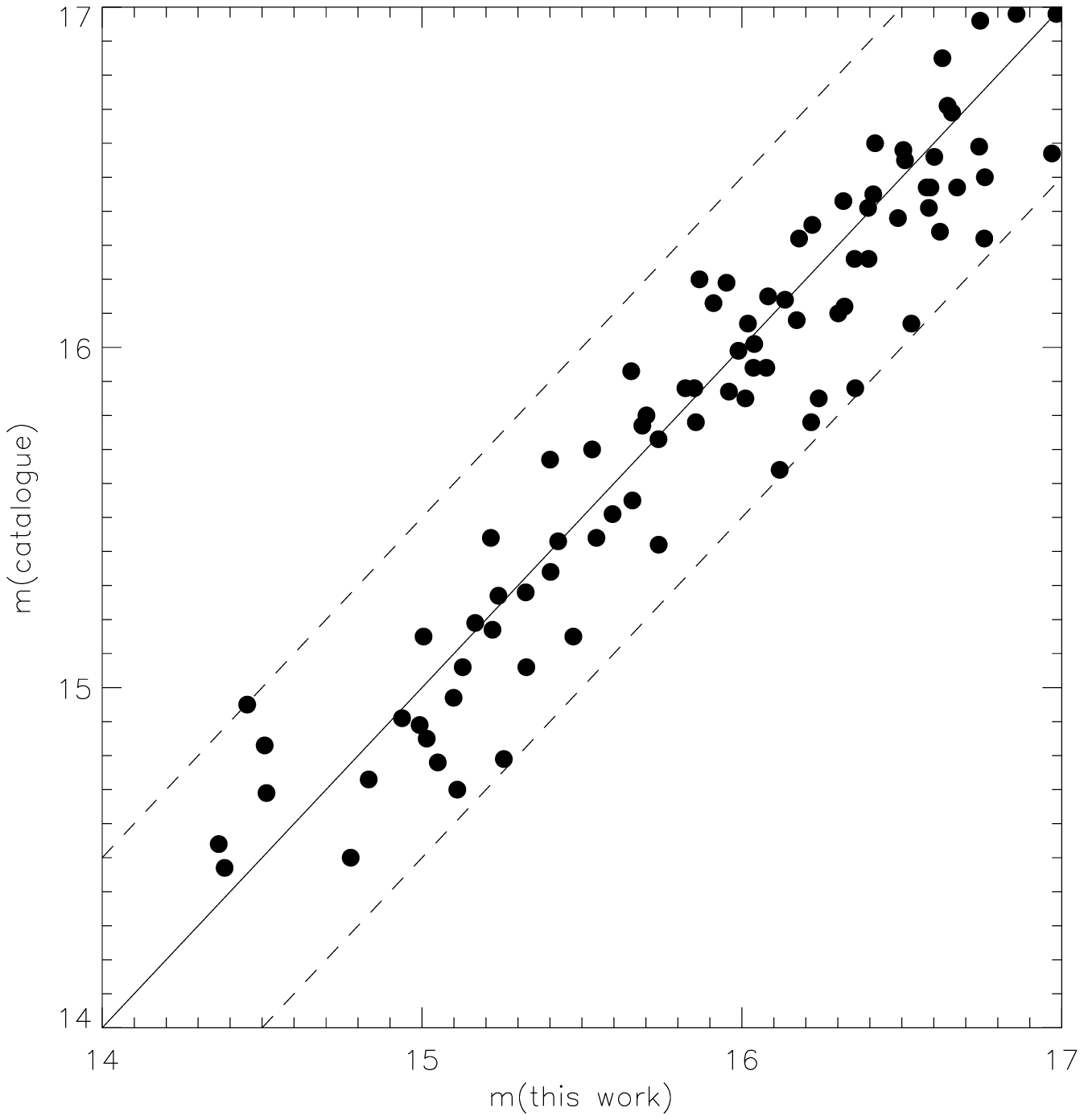}
\caption{Comparison between the magnitudes from Godwin et al. (1983) and  those
obtained from our fits. We have assumed in this plot a constant color difference of 0.6 magnitudes. The scatter of the points can be due, in part, to varying color differences for each galaxy.}
\end{figure}

\clearpage

\begin{figure}
\plotone{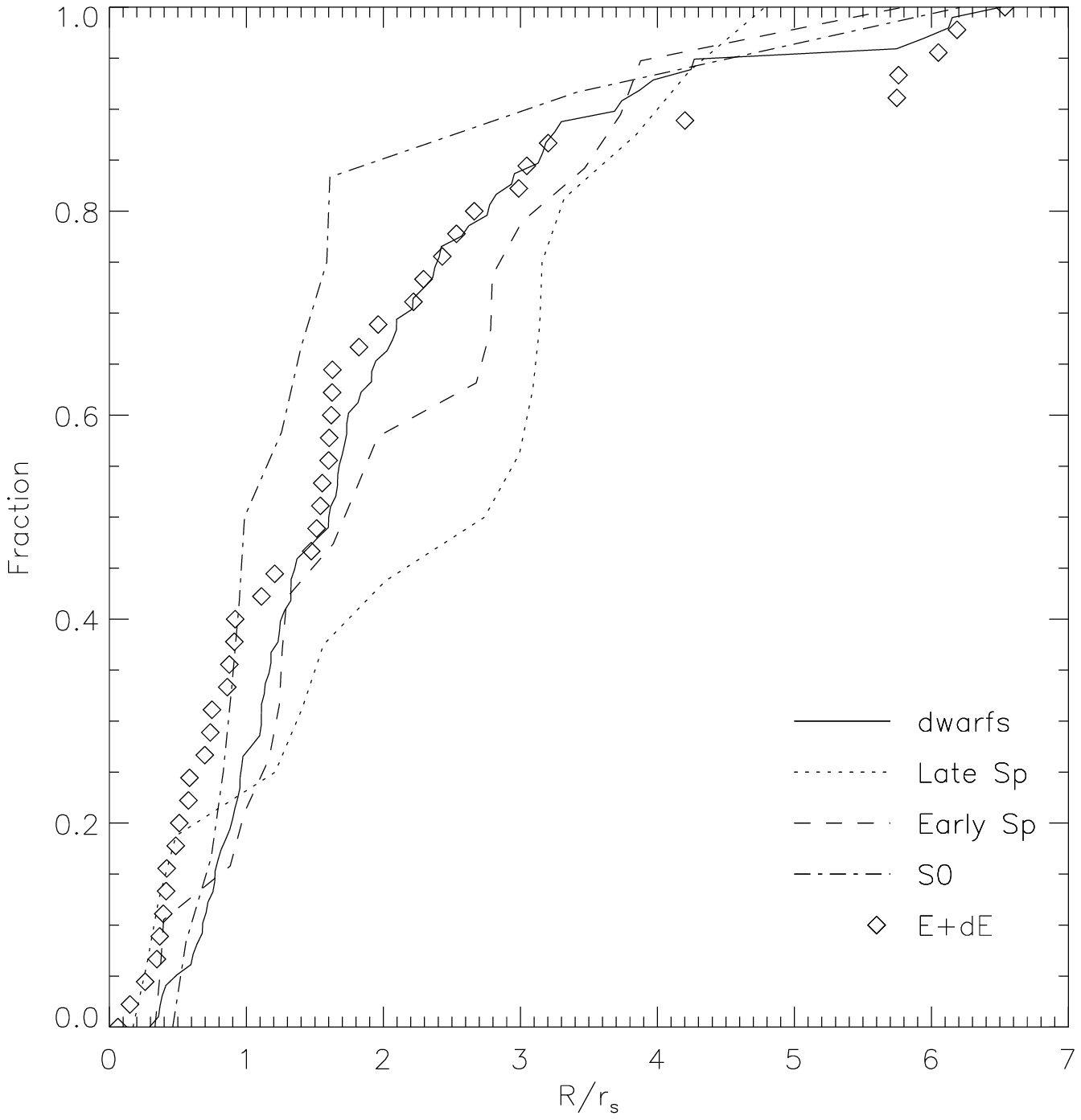}
\caption{Cumulative functions of the different morphological types as a function of the projected radius to the center of the cluster. \label{fig4}}
\end{figure}

\clearpage

\begin{figure}
\plotone{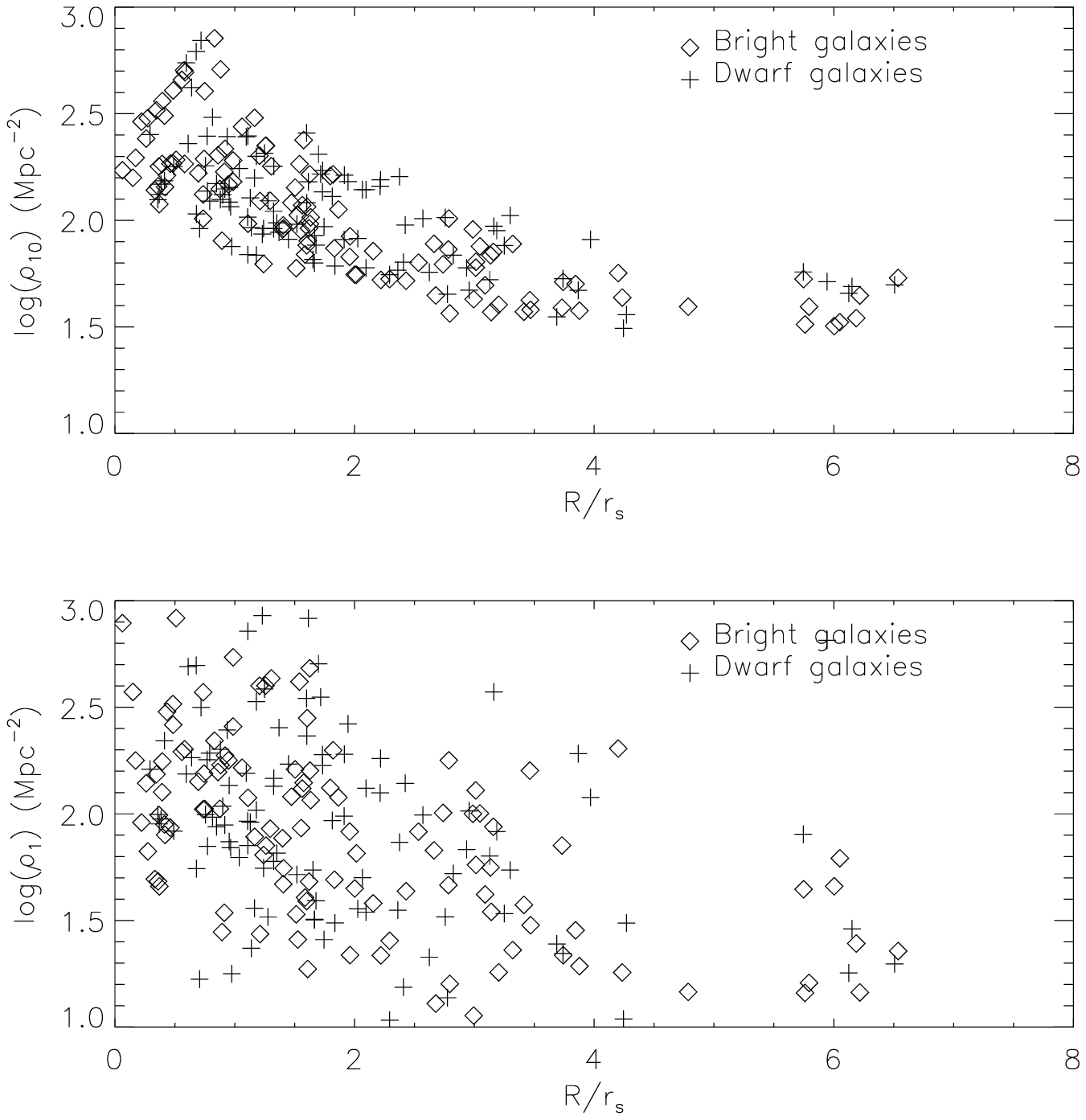}
\caption{ Projected density calculated with the 10 closest objects vs the projected radius ({\itshape top\/}).  Projected density computed taking  the nearest object into account ({\itshape bottom\/}). \label{fig5}}
\end{figure}

\clearpage

\begin{figure}
\plotone{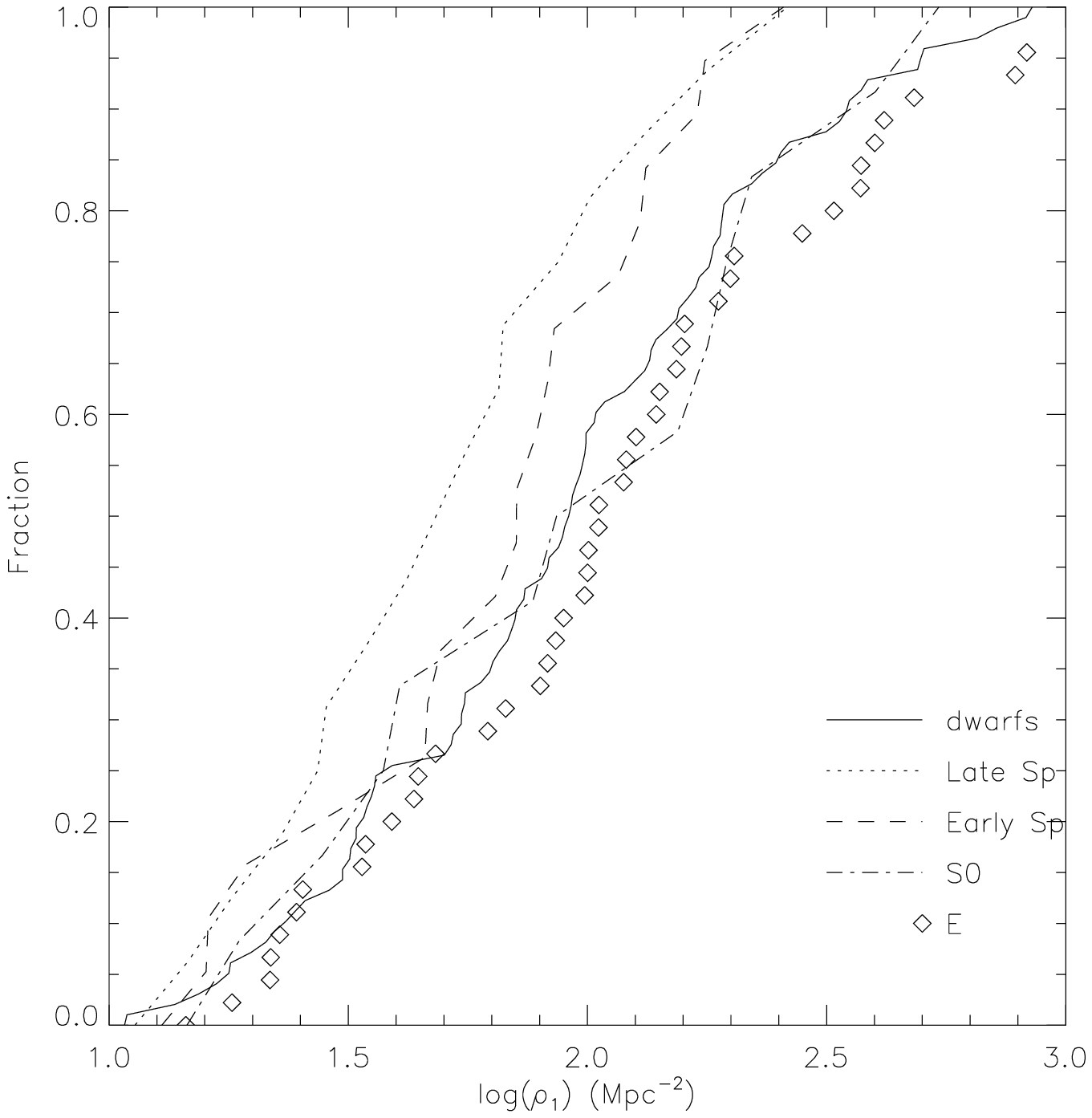}
\caption{Cumulative distribution functions of the different morphological types vs the projected density, $\rho_{1}$. See text for more details.\label{fig6}}
\end{figure}

\clearpage

\begin{figure}
\plotone{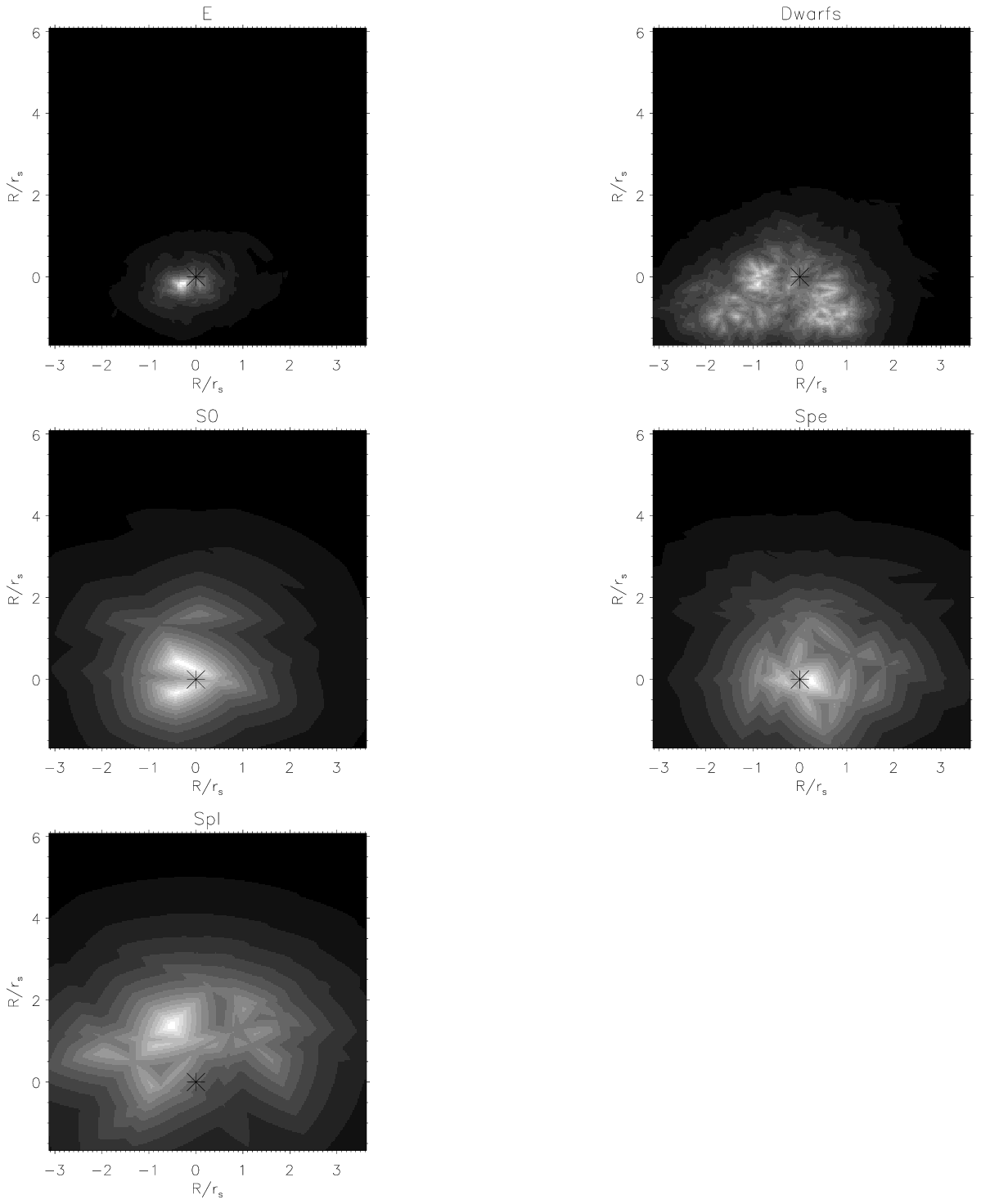}
\caption{Density images for the different types of galaxies, the black asterisk represent the center of the cluster (see text for details). In all panels
North is up and East is right.}
\end{figure}

\clearpage

\begin{figure}
\plotone{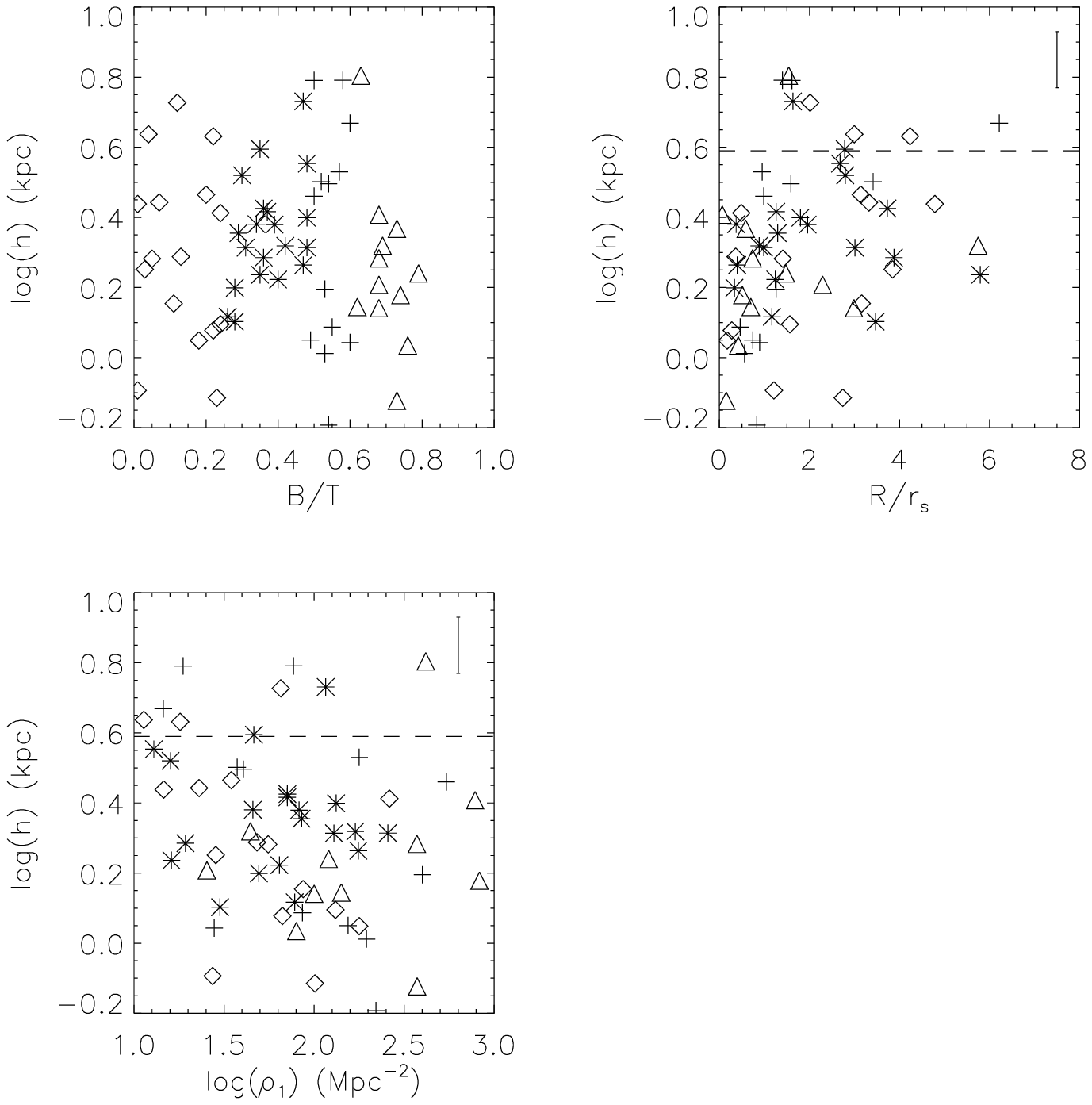}
\caption{Scale length of the disks in Coma vs. $B/T$ ({\itshape top left\/}), projected radius ({\itshape top right\/}), and projected density $\rho_{1}$ ({\itshape bottom\/}). The mean value of $\log(h)$ reported by Graham (2003) is overplotted (dashed line). The different symbols shows different 
morphological-type galaxies: Spl (diamonds), Spe (asterisks), S0 (crosses), and dE (triangles). Typical error bars in $\log(h)$ are overplotted in the panels. \label{fig7}}
\end{figure}

\clearpage

\begin{figure}
\plotone{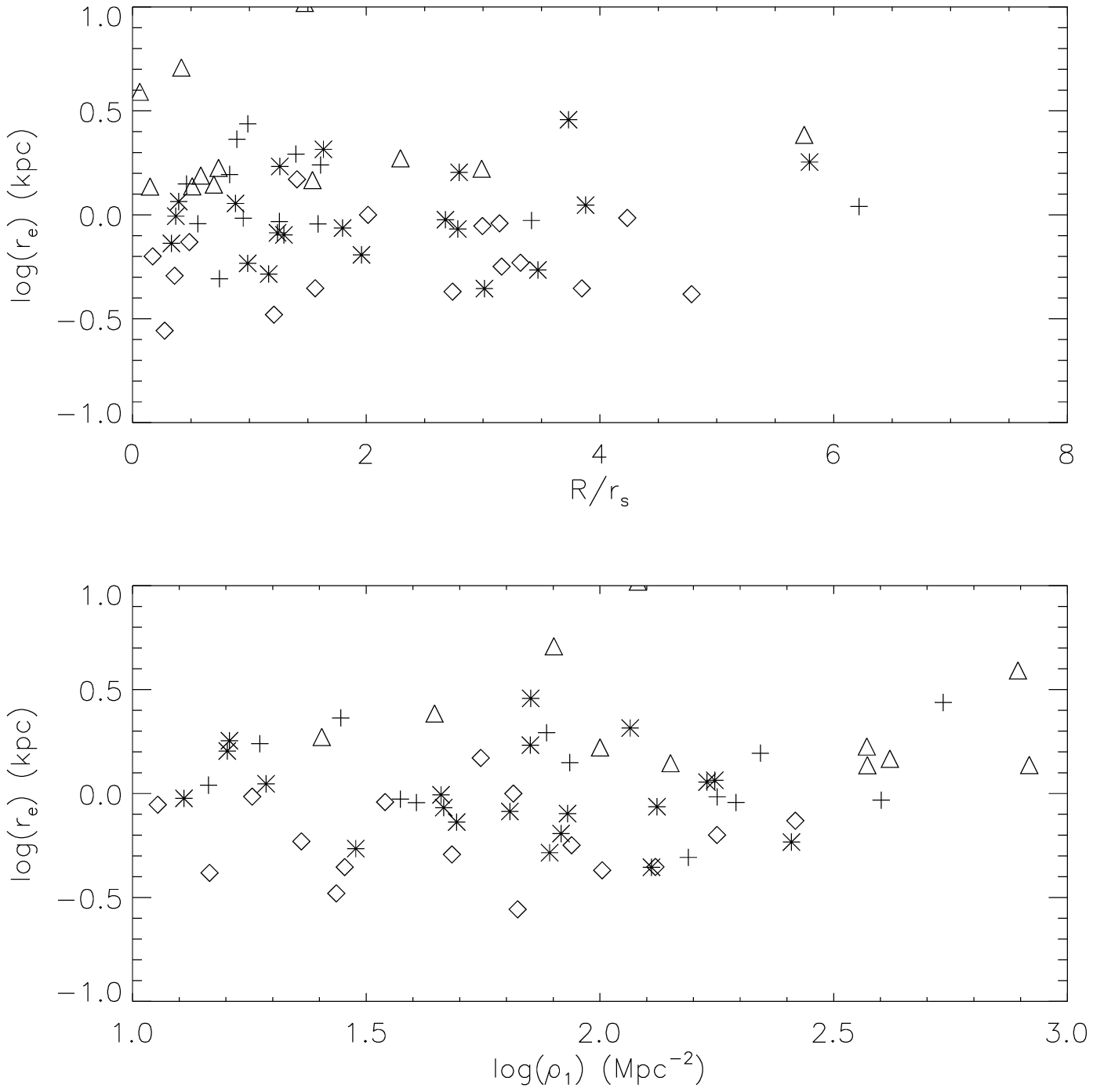}
\caption{Correlation of $r_{\rm e}$  vs.\ $\rho_{1}$ and $R/r_{\rm s}$. Symbols are as in Figure 11. \label{fig9}}
\end{figure}

\clearpage

\begin{figure}
\plotone{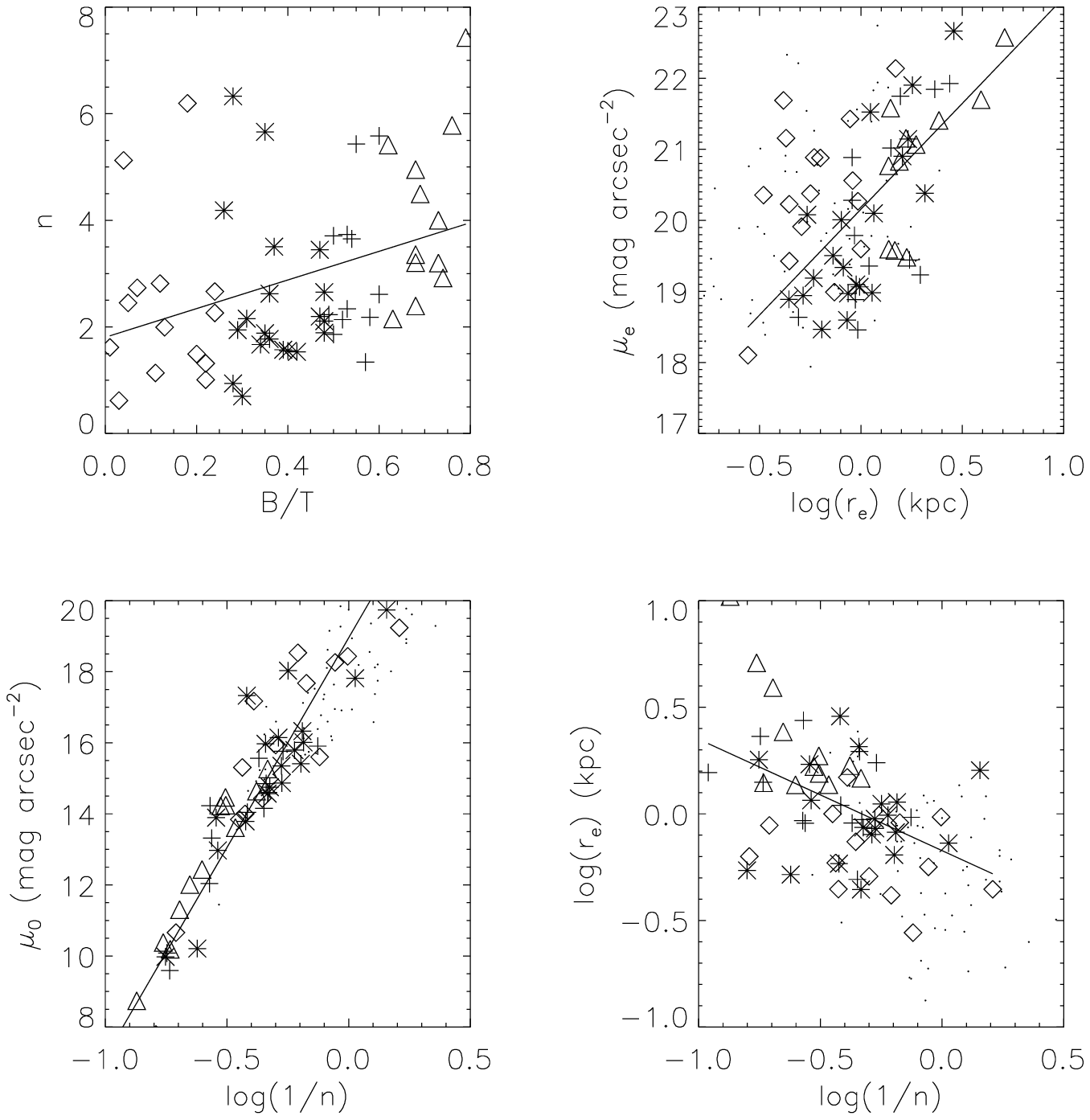}
\caption{Correlations of the bulge structural parameters $\mu_{\rm e}$, $r_{\rm e}$, and $n$. Symbols as in Figure 12. Small dots correspond to field galaxies (Graham 2001, 2003). \label{fig9}}
\end{figure}

\clearpage

\begin{figure}
\plotone{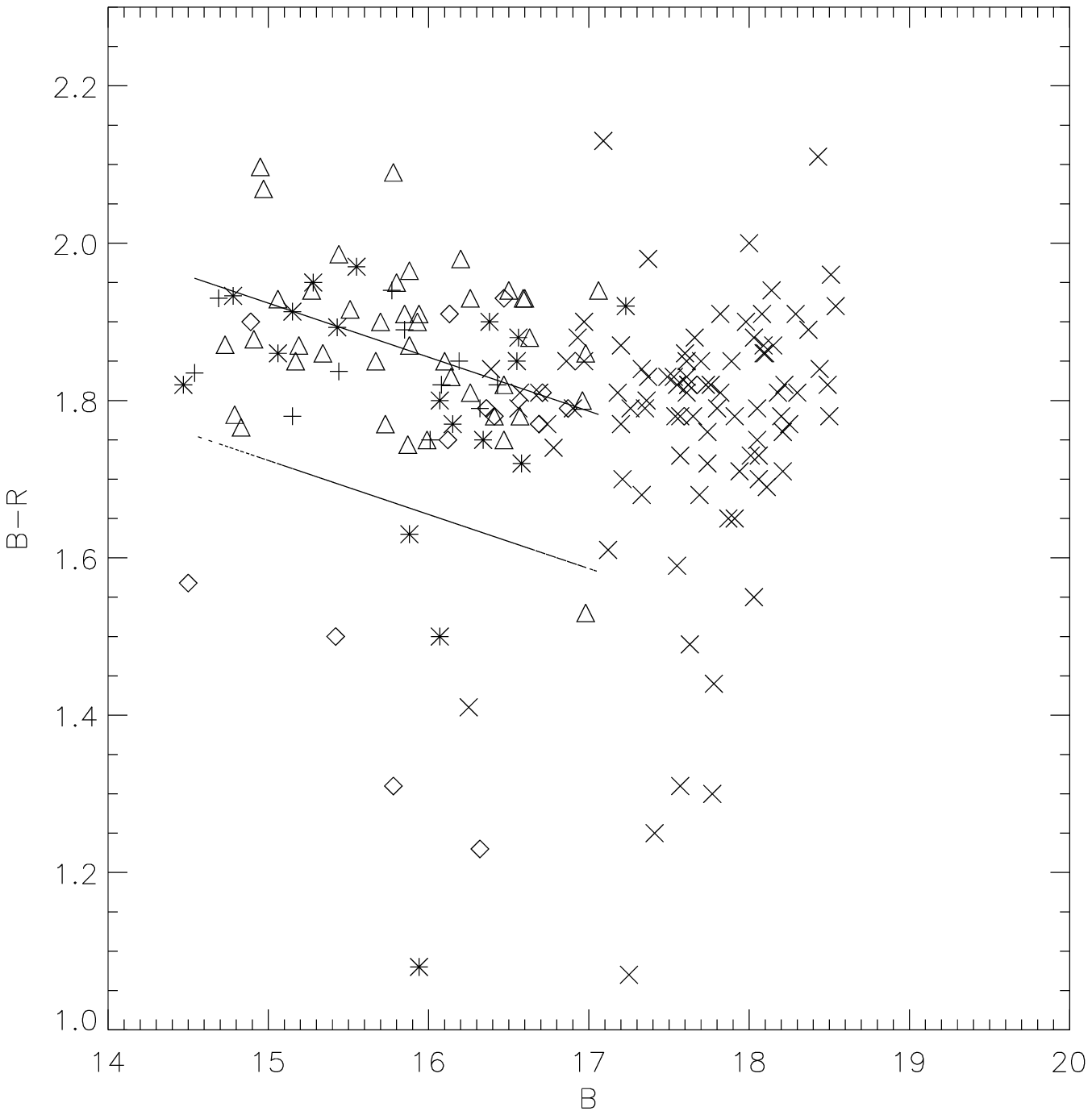}
\caption{Color--magnitude relation for galaxies with $m_{r}<17$ mag. The upper full line indicates the mean color of the E and S0 galaxies. The lower line represents the mean E/S0 color minus 0.2. Symbols are the same as in Figure 11. Crosses represents dwarf galaxies.}
\end{figure}

\clearpage

\begin{figure}
\plotone{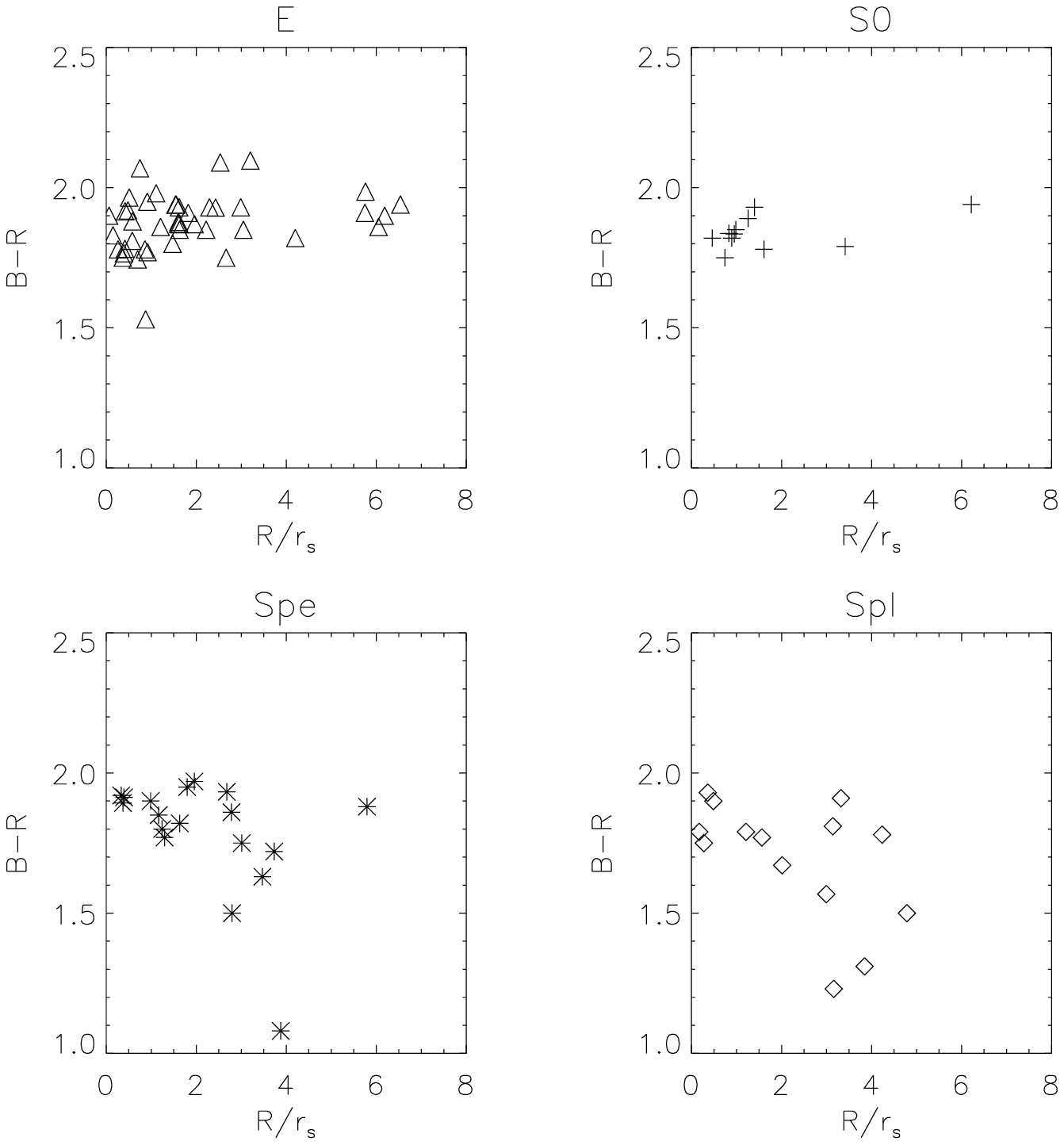}
\caption{$B-R$ colors vs.\ projected distance to the cluster center of the different types of galaxies.}
\end{figure}

\clearpage

\begin{figure}
\plotone{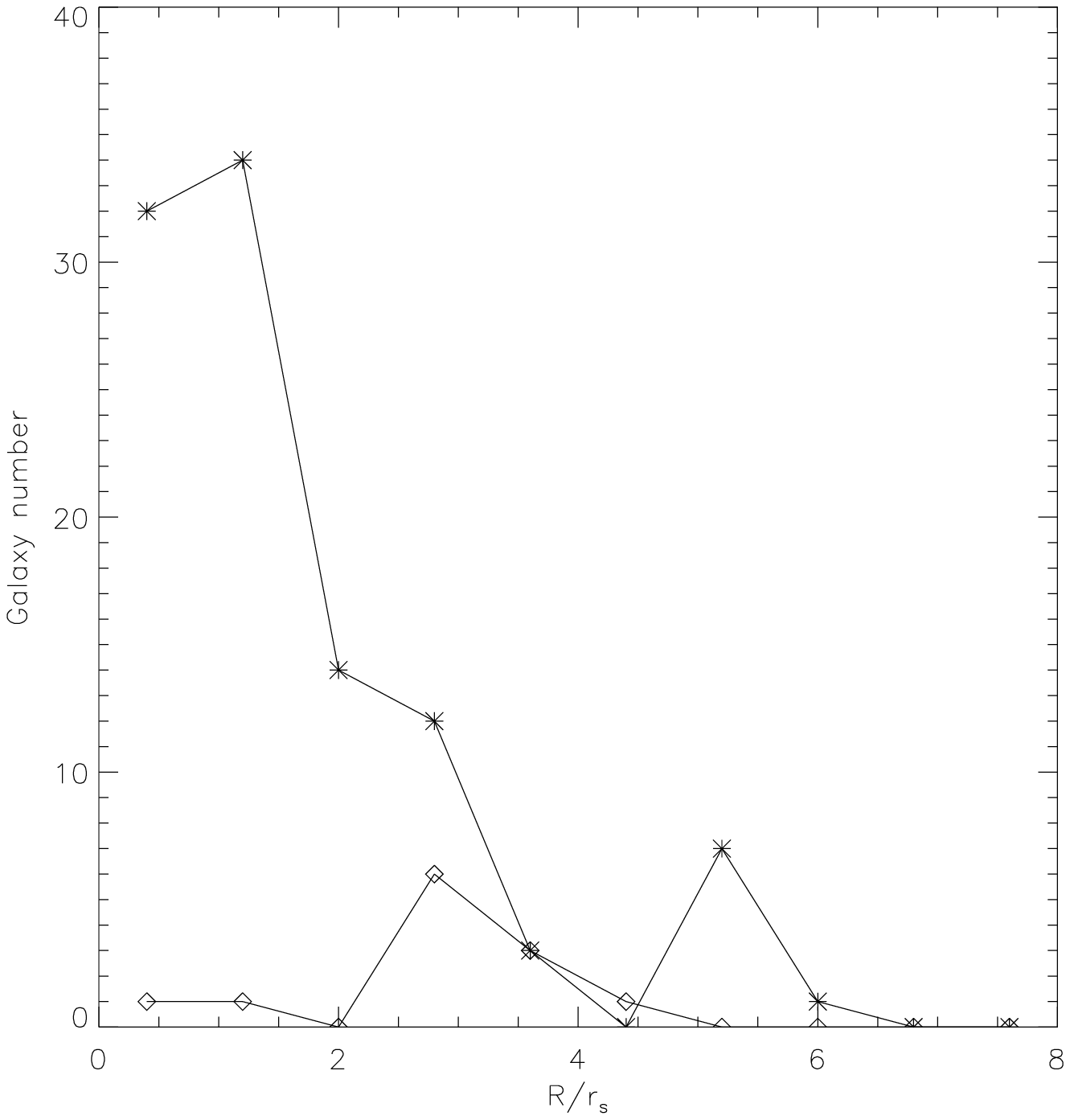}
\caption{Number of blue (diamonds) and red (asterisks) bright galaxies as a function of projected radius. }
\end{figure}

\clearpage

\begin{deluxetable}{ccccc}
\tabletypesize{\normalsize}
\tablecaption{Correlations of the scale length of the discs and the projected radius. \label{tbl-1}}
\tablewidth{0pt}
\tablehead{
\colhead{Galaxy type} & \colhead{Zero point}& \colhead{Slope} & \colhead{$r$} & \colhead{$P$}} 
\startdata
Spl & 0.14 $\pm$ 0.11  & 0.07 $\pm$ 0.04 & 0.41 & 0.01\\
Spe & 0.35 $\pm$ 0.06  & 0.03 $\pm$ 0.02 & 0.03 & 0.89\\
S0  & 0.17 $\pm$ 0.11  & 0.10 $\pm$ 0.05 & 0.50 & 0.07\\
Ed   & 0.22 $\pm $ 0.04  & 0.02 $\pm$ 0.04 & 0.16 & 0.57\\
All& 0.23 $\pm$ 0.05 & 0.05 $\pm$ 0.02 & 0.31 & 0.02\\
 \enddata
\end{deluxetable}

\begin{deluxetable}{ccccc}
\tabletypesize{\normalsize}
\tablecaption{Correlation of the scale length and the projected density $\rho_{1}$. \label{tbl-2}}
\tablewidth{0pt}
\tablehead{
\colhead{Galaxy type} & \colhead{Zero point}& \colhead{Slope} & \colhead{$r$} & \colhead{$P$}} 
\startdata
Spl & 0.78 $\pm$ 0.27  & $-$0.29 $\pm$ 0.15 & -0.44 & 0.07\\
Spe & 0.39 $\pm$ 0.18  & $-$0.02 $\pm$ 0.09 & -0.05 & 0.81\\
S0  & 0.90 $\pm$ 0.35  & $-$0.29 $\pm$ 0.17 & -0.45 & 0.10\\
dE   & 0.02 $\pm$ 0.25 & $-$0.09 $\pm$ 0.10 & -0.26 & 0.35 \\
All& 0.49 $\pm$ 0.11 & $-$0.09 $\pm$ 0.05 &-0.21 & 0.01 \\
\enddata
\end{deluxetable}

\end{document}